\documentclass[12pt]{article}
\usepackage{amsmath, amssymb, amsfonts, amsthm}
\usepackage{graphicx}

\usepackage{natbib}
\usepackage{url}
\usepackage[colorlinks,linkcolor=blue,citecolor=blue,urlcolor=blue]{hyperref}

\usepackage[utf8]{inputenc}
\usepackage[usenames,dvipsnames]{xcolor}
\usepackage{verbatim, color}
\usepackage{multirow}

\newtheorem{theorem}{\textsc{\bf Theorem}}
\newtheorem{definition}{\textsc{\bf Definition}}

\newtheorem{remark}{\textsc{\bf Remark}}

\newcommand{\bpm}{\begin{pmatrix}}
	\newcommand{\epm}{\end{pmatrix}}

\def\sign{\textup{sign}}
\def\NPN{{\textup{NPN}}}

\def\N{{\textup{N}}}

\def\E{\mathbb{E}}
\def\R{\mathbb{R}}

\def\bx{\mathbf{x}}
\def\bX{\mathbf{X}}
\def\bZ{\mathbf{Z}}
\def\bW{\mathbf{W}}
\def\by{\mathbf{y}}

\def\bSigma{\boldsymbol{\Sigma}}

\def\zeros{\boldsymbol{0}}

\defcitealias{Weinstein:2013jp}{TCGA, 2013}

\usepackage{algorithm}
\usepackage{enumitem}

\usepackage{authblk}
\usepackage{fullpage}

\title{\LARGE \bf Fast computation of latent correlations}

\author[1]{Grace Yoon}
\author[2]{Christian L. M\"uller}
\author[1]{Irina Gaynanova\footnote{Corresponding author. E-mail: irinag@stat.tamu.edu \\
		The authors gratefully acknowledge the support from the National Institutes of Health National Cancer Institute training grant T32-CA090301, the National Science Foundation grant DMS-1712943, and the Flatiron Institute of the Simons Foundation}}

\affil[1]{Department of Statistics, Texas A\&M University\\ 3143 TAMU, College Station, TX 77843}
\affil[2]{Center for Computational Mathematics, Flatiron Institute, New York, NY\\
	Department of Statistics, LMU M\"unchen, Munich, Germany\\
	
	Institute of Computational Biology, Helmholtz Zentrum M\"unchen, Germany}

\date{}

\begin{document}

		\maketitle
		
		\def\spacingset#1{\renewcommand{\baselinestretch}%
		{#1}\small\normalsize} \spacingset{1}
		
\begin{abstract}
Latent Gaussian copula models provide a powerful means to perform multi-view data integration 
since these models can seamlessly express dependencies between mixed variable types (binary,
continuous, zero-inflated) via latent \emph{Gaussian} correlations. The estimation of these latent
correlations, however, comes at considerable computational cost, having prevented the routine use of
these models on high-dimensional data. Here, we propose a new computational approach for
estimating latent correlations via a hybrid multi-linear interpolation and optimization scheme. 
Our approach speeds up the current state of the art computation by several orders of magnitude, thus allowing fast computation of latent Gaussian copula models even when the number of variables $p$ is
large. We provide theoretical guarantees for the approximation error of our numerical scheme and support its excellent performance 
on simulated and real-world data. We illustrate the practical advantages of our method on high-dimensional sparse quantitative and
relative abundance microbiome data as well as multi-view data from The Cancer Genome Atlas Project. Our method is implemented in the R package \textsf{mixedCCA}, available at \url{https://github.com/irinagain/mixedCCA}.
\end{abstract}

\noindent%
\vfill

\newpage
\spacingset{1.5} 

\section{Introduction}
\label{sec:intro}

Multi-view data, i.e, data collected on the same subjects from different sources or views, are
becoming increasingly common in the biomedical world thanks to advances in biological 
high-throughput technologies. Large-scale data collections, such as the Cancer 
Genome Atlas project \citepalias{Weinstein:2013jp}, make concurrent gene expression, 
methylation, mutation, and other data views with a mixed type of measurements (e.g., 
continuous, binary) readily available for multi-view data analysis. Moreover, recent 
sequencing-based technologies provide an abundance of high-dimensional biological data with 
excess zeros, ranging from Chip-Seq, to targeted amplicon and single-cell 
sequencing data. Many statistical analysis routines often start with estimating covariances 
and correlations from the different variables. 
However, standard Pearson sample covariance estimation via maximum likelihood estimation of covariance matrix 
is not well suited for these data since it is not able to handle the excess zeros in the data and its underlying normality 
assumption is violated by the highly skewed empirical data distributions.

Latent Gaussian copulas offer an elegant alternative for the analysis of multi-view data as they model 
associations between mixed variable types on the common latent Gaussian level, rather than on the mixed 
observed data level. \citet{Liu:2009wz} capture possible skewness in continuous measurements via Gaussian 
copula model. \citet{Fan:2016um} capture binary measurements via extra dichotomization step of Gaussian 
copulas, thus enabling joint modeling of continuous and binary variables. Extensions to ordinal variables have also been 
considered \citep{Quan:2018tv, Feng:2019wh}. \citet{mixedCCA, SPRING} capture variables with excess zeros 
via extra truncation step of Gaussian copula, thus enabling joint modeling of all continuous/binary/truncated (excess zeros) data
types. These models are very flexible and capture all dependencies via the common latent correlation matrix, which is estimated 
based on a robust rank-based measure of association (Kendall's $\tau$). By replacing Pearson sample correlation estimators with a 
rank-based correlation matrix estimator, latent Gaussian copula models have been shown to improve graphical model estimation 
\citep{Liu:2009wz, Fan:2016um, Feng:2019wh, SPRING}, canonical correlation analysis \citep{mixedCCA}, and discriminant analysis 
\citep{Han:2013ju}.
 
Despite the clear advantages offered by the latent Gaussian copula models, their widespread use on high-dimensional
biological data has been hindered by the considerable computational cost associated with 
the estimation of the latent correlation matrix $\bSigma$. Let $\sigma_{jk}$ be the latent correlation
between variables $j$ and $k$, and $\widehat \tau_{jk}$ be the corresponding sample Kendall's $\tau$. The two
are connected via the strictly increasing {bridge function $F$} such that $\E(\widehat \tau_{jk}) =
F(\sigma_{jk})$. This moment equation motivates the estimator $\widehat \sigma_{jk} = F^{-1}(\widehat \tau_{jk})$.
While the explicit form of $F$ has been derived for multiple variable types \citep{Fan:2016um, Quan:2018tv,
Feng:2019wh, mixedCCA}, its inverse $F^{-1}$ is not available in closed form. As a result, the estimation requires
solving a uniroot non-linear equation $F(x) = \widehat \tau_{jk}$ for every element of $\bSigma$. When the number of
variables $p$ is very large, this becomes computationally expensive. The computational cost also depends on the
type of variables (as it influences the form of $F$), and is especially problematic for truncated variable types, i.e., for
data with excess zeros such as single-cell and microbiome data. For instance, single-threaded computation of latent
correlations on a subset of the American Gut amplicon data \citep{McDonald2018} with $p=481$ species can take almost an hour
on a standard computer \citep{SPRING}. This makes repeated computations over sub-sampled or bootstrapped data or data
with thousands of variables computationally demanding.

\begin{figure}[!t]
\centering
\includegraphics[scale = 0.49, trim = .5cm 0 0 0cm, clip]{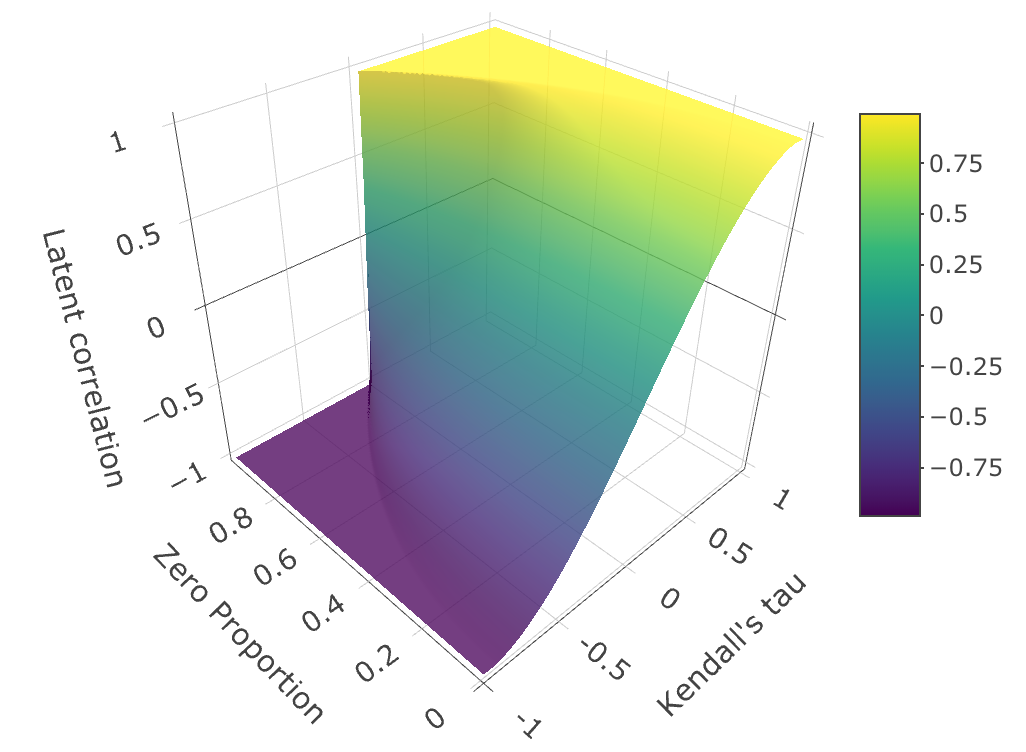}~~\includegraphics[width = 0.45\textwidth, trim = 0 0.5cm 0 0, clip]{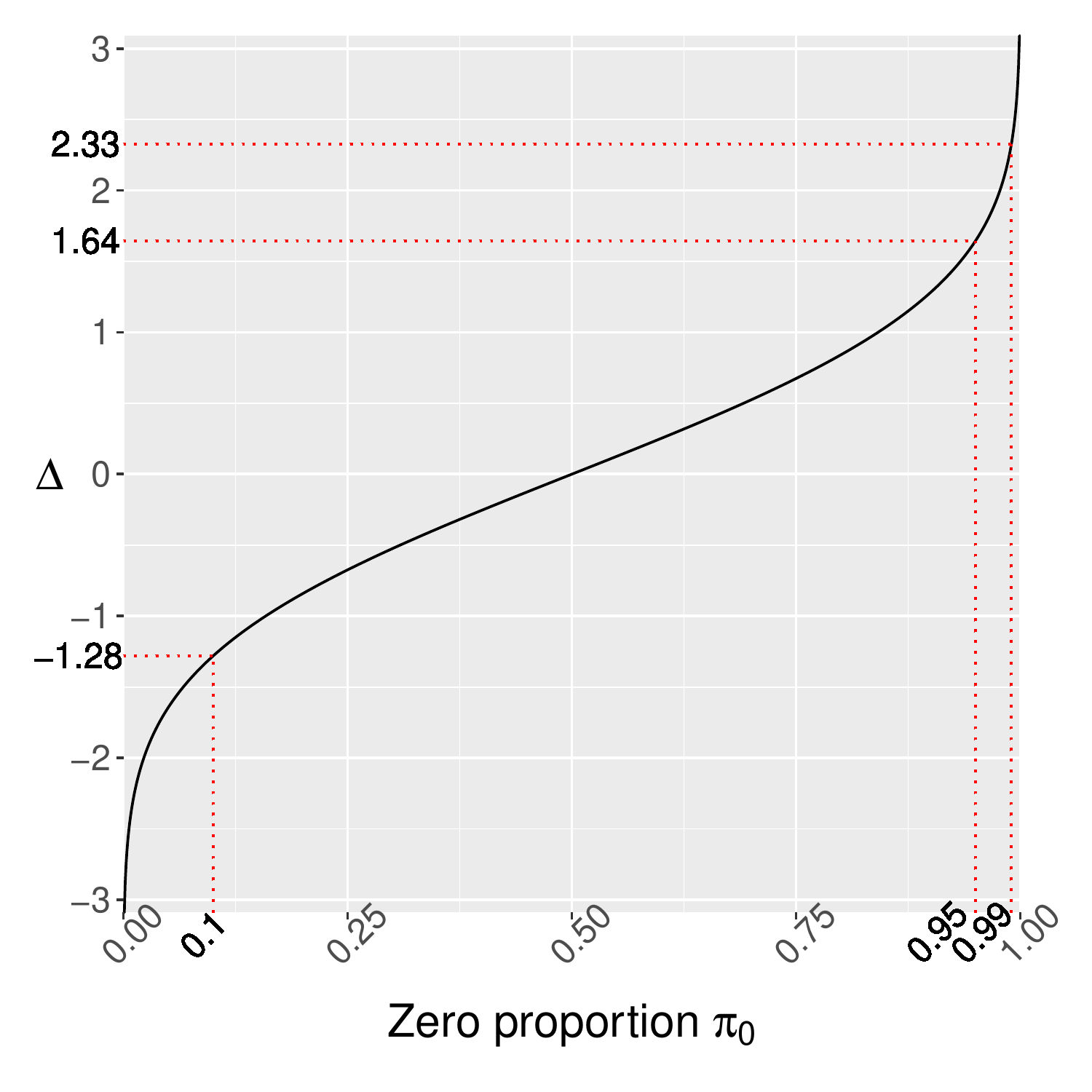}
\caption{(Left) Bridge inverse function $F^{-1}(\tau, \pi_0)$ for the continuous/truncated variables pair. The arguments are Kendall's $\tau$ (x-axis) and the proportion of zeros $\pi_0$ in the truncated variable (y-axis). The function values correspond to latent correlations (z-axis). (Right)~The estimated latent cutoff level $\Delta$ versus the proportion of zeros $\pi_0$ based on the moment equation $\widehat{\Delta} = \Phi^{-1}(\pi_{0})$. \label{fig:bridgeTCinv_3d}}
\end{figure}

Here, we overcome this challenge via a novel fast computation approach. Our idea is based on the observation that, even
though the exact analytic form of the inverse bridge function $F^{-1}$ is unknown, it is amenable to accurate multi-linear
interpolation of pre-computed function values over a well-chosen fixed grid of points. This pre-computation only needs to be
done once for each pair of variable types (continuous/binary/truncated), and is then readily available for any new dataset. Our 
interpolation scheme leads to dramatic reduction in computational cost (e.g., latent correlation on the 
American Gut microbiome data now only takes five minutes) while simultaneously controlling the approximation error required for
statistical estimation. To provide a visual illustration of the interpolation challenge, Figure~\ref{fig:bridgeTCinv_3d}
shows the surface of the inverse bridge function for the continuous/truncated variables pair, $F^{-1}(\tau, \pi_0)$, which
depends on the value of sample Kendall's $\tau$ and on the observed proportion of zeros $\pi_0$. While the function is
strictly increasing for each fixed value of $\pi_0$, its smoothness decreases significantly when $\pi_0$ increases. We present a 
hybrid interpolation scheme  that approximates the smooth part of the surface by multilinear interpolation of pre-computed 
function values over the fixed grid of point to obtain $F^{-1}(\tau_i, \pi_{0k})$, and explicit univariate non-linear optimization 
for the non-smooth part.

The rest of the paper is organized as follows. In Section~\ref{sec:model} we review the latent Gaussian copula model for mixed data
and the existing computational approach for latent correlation estimation. In Section~\ref{sec:interpolation} we propose a new fast
computation based on interpolation and provide theoretical guidance on the approximation error. In Section~\ref{sec:perfo}, we assess
the empirical performance both in terms of accuracy and speed on several high-throughput biological datasets.
Section~\ref{sec:conc} concludes with a discussion and future challenges. Our method is available in the R package
\textsf{mixedCCA} at \url{https://github.com/irinagain/mixedCCA}. A reproducible workflow of the presented numerical results is 
available at \url{https://github.com/GraceYoon/Fast-latent-correlation}.

\section{Latent correlation of latent Gaussian copula model}
\label{sec:model}

\subsection{Latent Gaussian copula model for mixed data}

We begin by reviewing the Gaussian copula model, or non-paranormal ($\NPN$) model, of \citet{Liu:2009wz} for possibly skewed
continuous data, e.g., gene expression. 

\begin{definition}[Continuous model]
A random $\mathbf{X}\in \R^p$ satisfies the Gaussian copula model if there exist monotonically increasing $f = (f_j)_{j=1}^p$ with $Z_j = f_j(X_j)$ satisfying $\bZ \sim \N_p(\zeros, \bSigma)$, $\sigma_{jj} = 1$;  $\bX \sim \NPN(\zeros, \bSigma, f)$.
\end{definition}
For binary data, such as mutation data, \cite{Fan:2016um} propose generalization of Gaussian copula via extra dichotomization step.
\begin{definition}[Binary model]
A random $\mathbf{X}\in \R^p$  satisfies the binary latent Gaussian copula model if there exists $\bW \sim \NPN(\zeros, \bSigma, f)$ such that $X_j = I(W_j > c_j)$, where $I(\cdot)$ is the indicator function and $c_j$ are constants.
\end{definition}
The binary model has been extended to ordinal variables with more than two levels \citep{Quan:2018tv, Feng:2019wh}.
For data with excess zeros, such as microbiome and single-cell data, \citet{mixedCCA} propose extra truncation of Gaussian copula.
\begin{definition}[Truncated model]
A random $\bX \in \R^p$ satisfies the truncated latent Gaussian copula model if there exists  $\bW \sim \NPN(\zeros, \bSigma, f)$ such that
$
X_j = I(W_j>c_j)W_j,
$
where $I(\cdot)$ is the indicator function and $c_j>0$ are constants.
\end{definition}
The \textit{mixed} latent Gaussian copula model jointly models $\mathbf{W}=(\mathbf{W}_1, \mathbf{W}_2, \mathbf{W}_3)\sim \NPN(\zeros, \bSigma, f)$ such that  $X_{1j} = W_{1j}$, $X_{2j} = I(W_{2j} > c_{2j})$ and $W_{3j} = I(W_{3j} > c_{3j})W_{3j}$.


\subsection{Bridge function}

The latent correlation matrix $\bSigma$ is the key parameter in the Gaussian copula models. Estimation of latent correlations is achieved via the bridge function $F$ such that $\E(\widehat\tau_{jk}) = F(\sigma_{jk})$, where $\sigma_{jk}$ is the latent correlation between variables $j$ and $k$, and  $\widehat \tau_{jk}$ is the corresponding sample Kendall's $\tau$. Given observed $\bx_j, \bx_k\in \R^n$,
\begin{equation}\label{eq:tau}
	\widehat \tau_{jk} = \widehat \tau(\bx_j,\bx_k) = \dfrac{2}{n(n-1)}\sum_{1\le i<i'\le n}\sign(x_{ij}-x_{i'j})\sign(x_{ik}-x_{i'k}),
\end{equation}
where $n$ is the sample size. Using $F$, one can construct $\widehat \sigma_{jk} = F^{-1}(\widehat \tau_{jk})$ with the corresponding estimator $\widehat \bSigma$ being consistent for $\bSigma$ \citep{Fan:2016um, Quan:2018tv, mixedCCA}. The explicit form of $F$ has been derived for all combinations of continuous(C)/binary(B)/truncated(T) variables \citep{Fan:2016um, mixedCCA}. We summarize these results below, and use CC, BC, TC, etc. to denote corresponding combinations.

\begin{theorem}\label{thm:bridge}
Let $\mathbf{W}_1 \in \R^{p_1}$, $\mathbf{W}_2 \in \R^{p_2}$, $\mathbf{W}_3 \in \R^{p_3}$ be such that $\mathbf{W}=(\mathbf{W}_1, \mathbf{W}_2, \mathbf{W}_3)\sim \NPN(\zeros, \bSigma, f)$ with $p = p_1 + p_2 + p_3$. Let $\mathbf{X}=(\mathbf{X}_1, \mathbf{X}_2, \mathbf{X}_3) \in \R^{p}$ satisfy $X_j = W_j$ for $j = 1, \ldots, p_1$, $X_j = I(W_j>c_j)$ for $j = p_1+1, \ldots, p_1+p_2$ and $X_j = I(W_j>c_j)W_j$ for $j = p_1+p_2+1, \ldots, p$ with $\Delta_j = f(c_j)$. The rank-based estimator of $\bSigma$ based on the observed $n$ realizations of $\bX$ is the matrix $\mathbf{ \widehat R}$ with $\widehat r_{jj}=1$, $\widehat r_{jk} = \widehat r_{kj}=F^{-1}(\widehat \tau_{jk})$ with block structure
\begin{equation}
    \begin{split}\nonumber
    \mathbf{ \widehat R} = \begin{pmatrix}
            F^{-1}_{\rm CC}(\widehat \tau) &  F^{-1}_{\rm CB}(\widehat \tau) & F^{-1}_{\rm CT}(\widehat \tau) \\
            F^{-1}_{\rm BC}(\widehat \tau)& F^{-1}_{\rm BB}(\widehat \tau)  & F^{-1}_{\rm BT}(\widehat \tau)\\
            F^{-1}_{\rm TC}(\widehat \tau)& F^{-1}_{\rm TB}(\widehat \tau)  & F^{-1}_{\rm TT}(\widehat \tau)
        \end{pmatrix}
    \end{split}
\end{equation}
\begin{equation}
    \begin{split}\nonumber
     F_{\rm CC}(r) & = \dfrac{2}{\pi} \sin^{-1}(r)\\
     F_{\rm BB}(r; \Delta_j, \Delta_k)&  = 2\left\{\Phi_2( \Delta_j,  \Delta_{k}; r)-\Phi( \Delta_j)\Phi( \Delta_{k})\right\}\\
     F_{\rm BC}(r; \Delta_j) & = 4\Phi_2( \Delta_j, 0; r/\sqrt{2})-2\Phi( \Delta_j)\\
     F_{\rm TB}(r;\Delta_j, \Delta_k) & =
    2 \{1-\Phi(\Delta_j)\}\Phi(\Delta_k)
    -2 \Phi_3\left(-\Delta_j, \Delta_k, 0; \bSigma_{3a}(r) \right)
    -2 \Phi_3\left(-\Delta_j, \Delta_k, 0; \bSigma_{3b}(r) \right)\\
    F_{\rm TC}(r;\Delta_j) & = -2 \Phi_2 (-\Delta_j,0; 1/\sqrt{2} ) +4\Phi_3 \left(-\Delta_j,0,0; \bSigma_3(r)\right)\\
      F_{\rm TT}(r;\Delta_j, \Delta_k)  & = -2 \Phi_4 (-\Delta_j, -\Delta_k, 0,0; \bSigma_{4a}(r)) + 2 \Phi_4 (-\Delta_j, -\Delta_k, 0,0; \bSigma_{4b} (r)),
    \end{split}
\end{equation}
with
\begin{equation}
    \begin{split}\nonumber
        \bSigma_{3a}(r)& =\bpm
1 & -r & 1/\sqrt{2} \\
-r & 1 & -r/\sqrt{2}\\
1/\sqrt{2} & -r/\sqrt{2} & 1
\epm, \quad
 \bSigma_{3b}(r)=\bpm
1 & 0 & -1/\sqrt{2} \\
0 & 1 & -r/\sqrt{2}\\
-1/\sqrt{2} & -r/\sqrt{2} & 1
\epm,\\
\bSigma_3(r) & = \bpm
        1 & 1/\sqrt{2} & r/\sqrt{2}\\
        1/\sqrt{2} & 1 & r\\
        r/\sqrt{2} & r & 1
    \epm, \quad
    \bSigma_{4a}(r) = \bpm
1 & 0 & 1/\sqrt{2} & -r/\sqrt{2}\\
0& 1 & -r/\sqrt{2}& 1/\sqrt{2}\\
1/\sqrt{2}& -r/\sqrt{2} & 1& -r\\
-r/\sqrt{2}& 1/\sqrt{2}& -r & 1
\epm\\
\bSigma_{4b}(r) & = \bpm
1 & r & 1/\sqrt{2} & r/\sqrt{2}\\
r & 1 & r/\sqrt{2}& 1/\sqrt{2}\\
1/\sqrt{2}& r/\sqrt{2} & 1& r\\ r/\sqrt{2}& 1/\sqrt{2}& r & 1
\epm.
    \end{split}
\end{equation}
Here $\Phi(\cdot)$ is the cdf of the standard normal distribution, and $\Phi_d(\cdot, \ldots, \cdot; \bSigma)$ is the cdf of the $d$-dimensional standard normal distribution with $d$-dimensional correlation matrix $\bSigma$. 
\end{theorem}

\subsection{Existing computation}\label{sec:orgcomp}

Theorem~\ref{thm:bridge} presents explicit forms of bridge functions for each data type combination. Using the selected bridge function,
the computation of latent correlation between two variables $j$ and $k$ is performed via Algorithm~\ref{alg:ORG}. Problem~\eqref{eq:opt}
has to be solved for all pairs of variables, leading to $O(p^2)$ computations. We refer to this approach as the original (ORG) computation
scheme.

\begin{algorithm}[!ht]
\caption{Original (ORG) method for latent correlation computation}\label{alg:ORG}
\textbf{Input:} $F(r) = F(r, \Delta_j, \Delta_k)$ - bridge function based on the type of variables $j$, $k$
\begin{enumerate}[leftmargin = *]
    \item Calculate $\widehat \tau_{jk}$ using \eqref{eq:tau}.
	\item For truncated/binary variable $j$, set $\widehat{\Delta}_j = \Phi^{-1}(\pi_{0j})$ with $\pi_{0j} = \sum_{i=1}^{n} I(x_{ij} = 0)/n$.
	\item Compute $F^{-1} (\widehat \tau_{jk})$ as 
	\begin{equation}\label{eq:opt}
	    \widehat r_{jk} = \arg\min_{r} \left\{F(r) - \widehat\tau_{jk}\right\}^2,
	\end{equation}
	where \eqref{eq:opt} is solved via \textsf{optimize} function in \textsf{R}.
\end{enumerate}
\end{algorithm}


\section{Inversion via multilinear interpolation}\label{sec:interpolation}

The inverse bridge function is an analytic function of at most three parameters: (i) Kendall's $\tau$, (ii) proportion of zeros in the 1st variable and (possibly) (iii) proportion of zeros in the 2nd variable (see Theorem~\ref{thm:bridge}). We propose to pre-calculate the function on a fixed 2d (or 3d) grid, and perform multilinear interpolation to estimate its values on a new set of arguments.

\subsection{Multilinear interpolation}
\label{sec:ml}


\begin{definition}[Bilinear interpolation] Suppose we have 4 neighboring data points $f_{ij} = f(x_i, y_j)$ at $(x_i, y_j)$ for $i, j\in \{0, 1\}$. For $\left\{ (x, y) | x_0 \le x \le x_1, y_0 \le y \le y_1 \right\}$,   
the bilinear interpolation at $(x, y)$ is
\begin{equation}
    \begin{split}
        \widetilde f(x, y) & = (1-\alpha)(1-\beta)f_{00} + (1-\alpha)\beta f_{01} +  \alpha(1-\beta) f_{10} + \alpha\beta f_{11}
    \end{split}
\end{equation}
where $\alpha = (x - x_0)/(x_1 - x_0)$ and $\beta = (y - y_0)/(y_1 - y_0)$.
\end{definition}

\begin{definition}[Trilinear interpolation] Suppose we have 8 neighboring data points $f_{ijk} = f(x_i, y_j, z_k)$ at $(x_i, y_j, z_k)$ for $i, j, k \in \{0, 1\}$. For $\left\{ (x, y, z) | x_0 \le x \le x_1, y_0 \le y \le y_1, z_0 \le z \le z_1 \right\}$,
the trilinear interpolation at $(x, y, z)$ is
\begin{equation}
    \begin{split}
        \widetilde f(x, y, z) & = (1-\alpha)(1-\beta)(1-\gamma) f_{000} + (1-\alpha)(1-\beta)\gamma f_{001}
        + (1-\alpha)\beta(1-\gamma) f_{010}\\
        & \quad + \alpha(1-\beta)(1-\gamma) f_{100}
        + (1-\alpha)\beta\gamma f_{011} + \alpha(1-\beta)\gamma f_{101}
        + \alpha\beta(1-\gamma) f_{110}\\
         & \quad + \alpha \beta \gamma f_{111}
    \end{split}
\end{equation}
where $\alpha = (x - x_0)/(x_1 - x_0)$, $\beta = (y - y_0)/(y_1 - y_0)$ and $\gamma = (z - z_0)/(z_1 - z_0)$.
\begin{figure}
\centering
\includegraphics[width = 2.9in, height = 2.6in, trim = 4cm 4cm 7cm 2cm, clip]{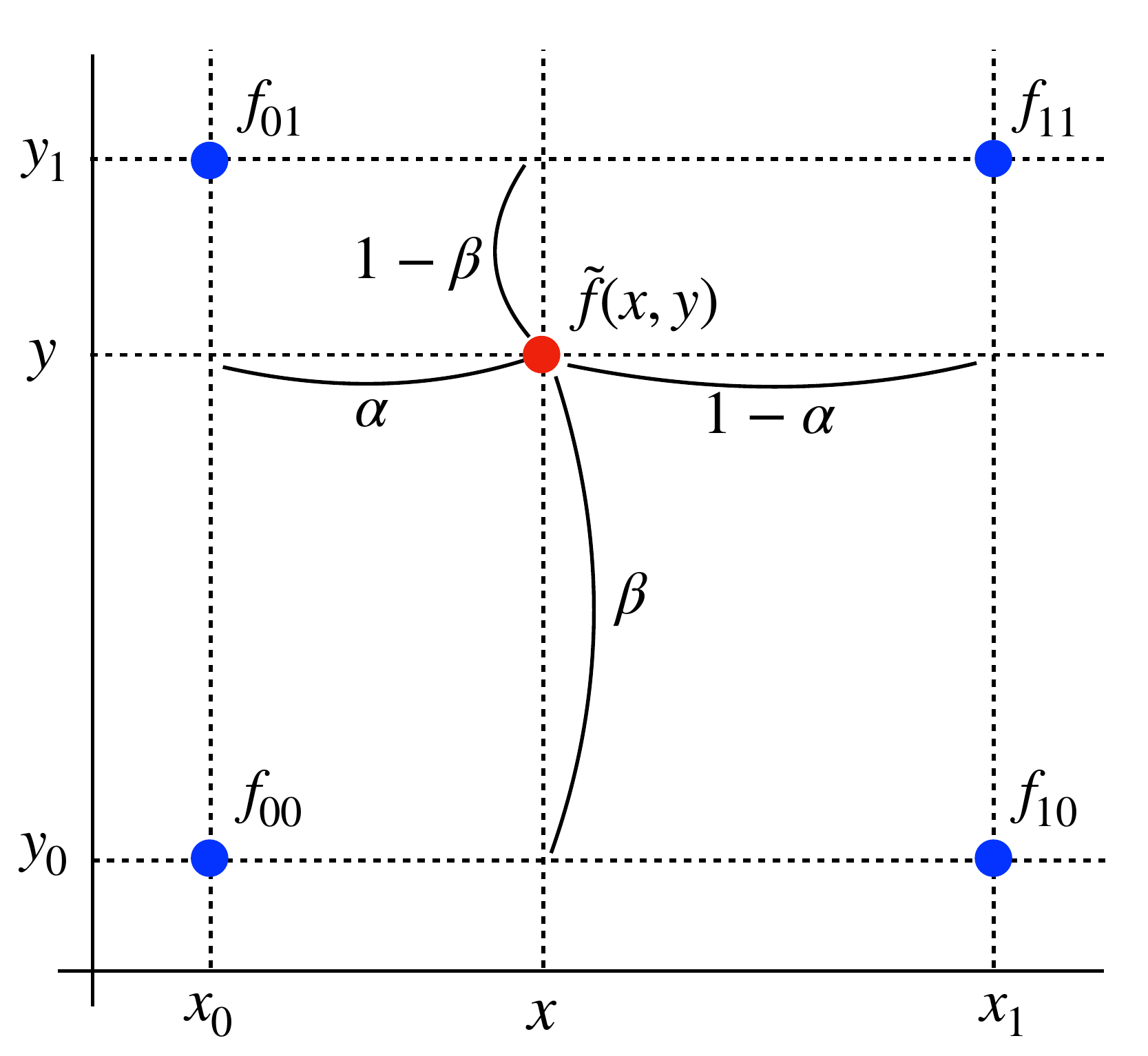}
\quad 
\includegraphics[width = 3.1in, height = 2.6in, trim = 4.5cm 4.6cm 4.1cm 1.4cm, clip]{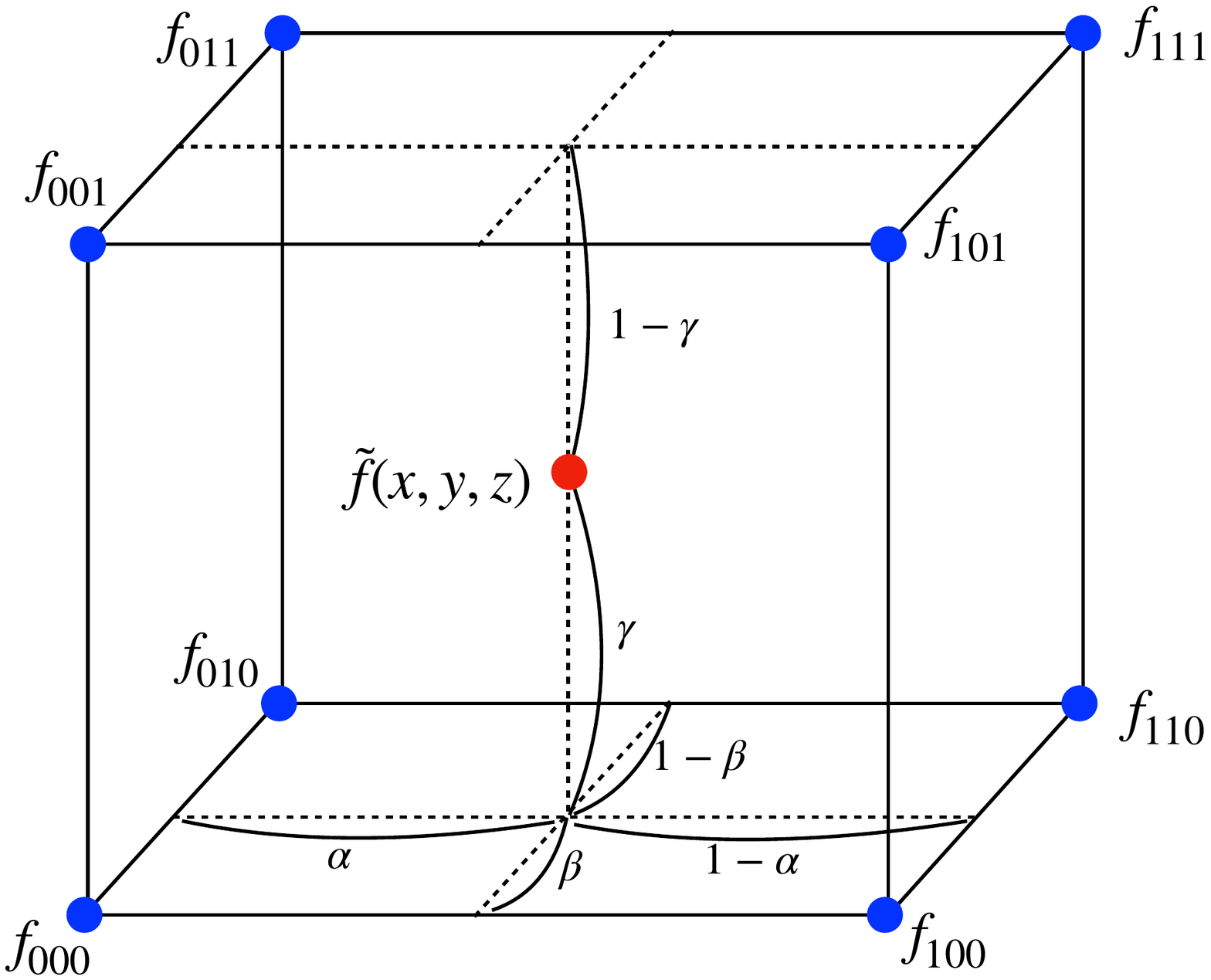}
\caption{Bilinear (Left) and trilinear (Right) interpolation.} \label{fig:ipol_3d}
\end{figure}

\end{definition}

In short, $d$-dimensional multilinear interpolation uses a weighted average of $2^d$ neighbors to approximate the function values at the points within the $d$-dimensional cube of the neighbors, see Figure~\ref{fig:ipol_3d}.

\subsection{Error bound for multilinear interpolation}\label{sec:errbd}
\citet{Alan1988} provide an error bound for multilinear interpolation.

\begin{theorem}\label{thm:errbd}
For a function $f:\mathbb{R}^d \to \mathbb{R}$, assume that the function values are given at $2^d$ points $ f(x_{1i}, \ldots, x_{di})
\text{ for } i = 0, 1$. Let $\widetilde f:\mathbb{R}^d \to \mathbb{R}$ denote the multilinear interpolation function of $f$ on the
$d$-dimensional cube $\Omega = \left\{(x_1, \ldots, x_d): x_{10} < x_1 < x_{11}, \ldots, x_{d0} < x_d < x_{d1}\right\}$ using the given
$2^d$ neighboring points. Then, for every point $\mathbf{x} = (x_1, \ldots, x_d)^\top \in \Omega$
\begin{equation}
    \begin{split}\label{eq:errbd}
        |f(\mathbf{x}) - \widetilde f(\mathbf{x})| & \le \dfrac{d}{8}h^2 \sup_{\substack{i=1, \ldots,d}} \left|\dfrac{\partial^2 f(\mathbf{x})}{\partial x_i^2}\right|,
    \end{split}
\end{equation}
where $h = \max_{j = 1,\ldots, d}\left| x_{j1} - x_{j0}\right|$. 
\end{theorem}

Theorem~\ref{thm:errbd} shows that the error bound in our proposed approximation via multilinear interpolation depends on the second
derivative of the bridge inverse function. The dimension $d=2$ for the BC and TC cases, and $d = 3$ for the TT, TB, and
BB cases. While the bridge inverse functions are differentiable, the explicit forms of derivatives are difficult to calculate
analytically. Nevertheless, we were able to derive explicit bounds for the BC and the TC case, respectively,  thus providing
theoretical guidance on the aspects of the models that affect interpolation accuracy. The proofs 
are in
the Supplementary Materials.

\begin{theorem}\label{thm:BCbound}
Let $F^{-1}(\tau, \Delta)$ be the inverse bridge function for the binary/continuous case. 
Let $\Delta$ satisfy $|\Delta| \le M$ for some constant $M$. Then
\begin{equation}
    \begin{split}\label{eq:errbd-bc}
        |F^{-1}(\tau, \Delta) -\widetilde F^{-1}(\tau, \Delta)| & \le 2h^2 |F^{-1}(\tau, \Delta)| (2M^2 + 1) \exp(M^2),
    \end{split}
\end{equation}
where $h$ is the maximal grid width. 
\end{theorem}
Theorem~\ref{thm:BCbound} shows that the approximation error in the BC case strongly depends on the absolute size of $\Delta$. Since
we estimate $\Delta$ as $\Phi^{-1}(\pi_0)$ (Algorithm~\ref{alg:ORG}), and $\pi_0$ is the observed proportion of zeros,
Theorem~\ref{thm:BCbound} implies that the approximation is more accurate when the numbers of zeros and ones are balanced ($\Delta
\approx 0$), and less accurate when they are unbalanced (see left panel in Figure~\ref{fig:bridgeTCinv_3d} for the correspondence 
between $\Delta$ and $\pi_0$). The dependence on the latent correlation $r=F^{-1}(\tau, \Delta)$ is less strong. Nonetheless, the 
accuracy decreases as $|r|$ increases. 
%
%
%
\begin{theorem}\label{thm:TCbound}
Let $F^{-1}(\tau, \Delta)$ be the inverse bridge function for the truncated/continuous case. Let $\Delta$ be such that $\Delta \leq M$ for some positive constant $M$. Then
\begin{equation}
    \begin{split}\label{eq:errbd-tc}
        |F^{-1}(\tau, \Delta) -\widetilde F^{-1}(\tau, \Delta)| & \le  \dfrac{4h^2}{\left\{\Phi(-\sqrt{2}M)\right\}^2} \max \left( \dfrac{|F^{-1}(\tau, \Delta)|}{\Phi(-\sqrt{2}M)},   \sqrt{1-\left\{F^{-1}(\tau, \Delta)\right\}^2} \right),
    \end{split}
\end{equation}
where $h$ is the maximal grid width. 
\end{theorem}
Theorem~\ref{thm:TCbound} shows that the approximation error in the TC case strongly depends on how large is $\Delta$. This is similar to the BC case. However, in the TC case, $\Delta$ only needs to be bounded from above. This is because as $\Delta$ goes to negative infinity, the truncated data type gets closer to the continuous one as the proportion of zeros $\pi_0$ goes to zero (see left panel in Figure~\ref{fig:bridgeTCinv_3d}). On the other hand, as $M$ increases, $\Phi(-\sqrt{2}M)$ goes to 0 making the upper bound in Theorem~\ref{thm:TCbound} very large. For example, if $M = 1.64$ (95\% zeros, see Figure~\ref{fig:bridgeTCinv_3d}), then 
$1/\Phi(-\sqrt{2}M)^3 \approx 945099$. The size of the latent correlation has a milder effect on accuracy. Nonetheless, the accuracy decreases as $|r| = |F^{-1}(\tau, \Delta)|$ increases.

In summary, the approximation accuracy of our approach is affected by the observed proportion of zeros (through the size of $M$) 
and by the size of latent correlation (the actual function value at the interpolation point). The interpolation accuracy is poor for
binary data when the numbers of zeros and ones are extremely unbalanced, and for truncated data, when the proportion of zero values is
close to 1.

\begin{remark} The estimation consistency of the original method (Algorithm~\ref{alg:ORG}) is established under conditions that all
correlation values are bounded away from one, and that the values of $\Delta$ are bounded \citep{Fan:2016um,mixedCCA}. 
Theorems~\ref{thm:BCbound}--\ref{thm:TCbound} reveal that the same conditions are required for good interpolation approximation,
thus emphasizing a close connection between statistical (estimation) and computational (approximation) accuracy.
\end{remark}

%
%

\subsection{Numerical implementation}\label{sec:hybridcomp}

Algorithm~\ref{alg:ML} summarizes the multilinear interpolation approach. 
\begin{algorithm}
\caption{Multilinear Interpolation (ML) method for latent correlation computation}\label{alg:ML}
\textbf{Input:} Pre-computed values $F^{-1}(\tau_l, \Delta_m, \Delta_q)$ on a fixed grid $(\tau_{l}, \Delta_m, \Delta_q) \in \mathcal{G}$ based on the type of variables $j$ and $k$.
\begin{enumerate}
    \item[1-2.] Same as Algorithm~\ref{alg:ORG}.
	\item[3.] Set $\widehat r_{jk} = \widetilde{F}^{-1}(\widehat \tau_{jk}, \widehat \Delta_j, \widehat \Delta_k)$, where  $\widetilde F^{-1}$ is the trilinear interpolation of $F^{-1}$ using $\mathcal{G}$.
\end{enumerate}
\end{algorithm}
%
%

We next present a hybrid scheme to prevent interpolation in regions with high approximation errors. From Theorems~\ref{thm:BCbound}
and \ref{thm:TCbound}, the approximation error increases when (i) the proportion of zeros $\pi_0$ increases and (ii) the absolute
value of latent correlation $r$ is large, i.e., large absolute values of Kendall's $\tau$. However, the range
of $\widehat \tau$ values is directly affected by $\pi_0$ since $\sign(x_{ij}-x_{i'j})\sign(x_{ik}-x_{i'k}) = 0$ in~\eqref{eq:tau} for all pairs $(i, i')$ with zero values. That is, higher $\pi_0$ leads to smaller range of $\widehat \tau$. We derive upper bounds on the values of $\widehat \tau$ as a function of $\pi_0$ and use these bounds to define the boundary region for interpolation. 

Let $\bx \in \R^n$ and $\by \in \R^n$ be the observed $n$ realizations of truncated continuous and continuous variable, respectively. The upper bound on the range of Kendall's $\tau$ can be obtained by enumerating the number of pairs between zero values. Let $\pi_{0} = n_0/n$ where $n_{0} = \sum_{i=1}^{n} I(x_{i}=0)$ is the number of zero values out of $n$. Then from~\eqref{eq:tau},
\begin{equation}
    \begin{split}
        	| \widehat\tau(\bx,\by) | \le \left\{ {n \choose 2} - {n_{0} \choose 2}  \right\} \Big/ {n \choose 2} \le 1 - \dfrac{n_{0}(n_{0} - 1)}{n(n - 1)} \approx 1 - \pi_0^2.
    \end{split}
\end{equation}
Similarly, we can approximate the range of Kendall's $\tau$ for other data type combinations (see the
Supplementary Materials). In summary, we obtain the following approximate bound (ABD) on the range of $|\widehat \tau|$ values
%
\begin{equation}\label{eq:boundary}
\text{ABD} =
\begin{cases}
      1 - \left(\pi_0\right)^2 & \text{ for TC case} \\
     1 - \left\{ \max(\pi_{0x}, \pi_{0y}) \right\}^2   & \text{ for TT case} \\
    2\pi_0 \left( 1- \pi_0 \right)  & \text{ for BC case} \\
    2\min(\pi_{0x}, \pi_{0y}) \left\{ 1- \max(\pi_{0x}, \pi_{0y})\right\}  & \text{ for BB case} \\
  {2\max(\pi_{0y}, 1 - \pi_{0y}) \{1 - \max(\pi_{0y}, 1 - \pi_{0y}, \pi_{0x})\}} & \text{ for TB case} \\
\end{cases}
\end{equation}

{ If value of $|\widehat \tau|$ is close to \text{ABD}, this indicates a high value of zero proportion and a high value of correlation. To prevent high approximation errors, we propose to apply linear interpolation if $|\widehat \tau|\leq 0.9\text{ABD}$, and to use the original estimation approach otherwise.}
%
%
%
We call this the hybrid multilinear interpolation with boundary (MLBD) algorithm (Algorithm~\ref{alg:MLBD}).
\begin{algorithm}
\caption{Multi-Linear interpolation with Boundary (MLBD) method}\label{alg:MLBD}
\textbf{Input:} Pre-computed values $F^{-1}(\tau_l, \Delta_m, \Delta_q)$ on a fixed grid $(\tau_{l}, \Delta_m, \Delta_q) \in \mathcal{G}$ based on the type of variables $j$ and $k$.
\begin{enumerate}
    \item[1-2.] Same as Algorithm~\ref{alg:ORG}.
	\item[3.] If $|\widehat{\tau}_{jk}| \le 0.9{\rm ABD}$ in~\eqref{eq:boundary}, apply ML Algorithm~\ref{alg:ML}.\\
	If $|\widehat{\tau}_{jk}| > 0.9{\rm ABD}$, apply ORG Algorithm~\ref{alg:ORG}.
\end{enumerate}
\end{algorithm}



We use the same grid for both Algorithms~\ref{alg:ML} and~\ref{alg:MLBD}, it is implemented in the R package~\textsf{mixedCCA}. The detailed description of the grid is in the Supplementary Materials.


\section{Performance Assessment}\label{sec:perfo}
We assess the approximation quality and computational speed of three algorithm for latent correlation estimation: the ORG method summarized in Algorithm~\ref{alg:ORG}, the multilinear interpolation scheme (ML) in Algorithm~\ref{alg:ML}, and the hybrid MLBD scheme in Algorithm~\ref{alg:MLBD}. 

\subsection{Approximation accuracy of latent correlation estimation}\label{sec:accuracy}
We first focus on the approximation accuracy in computing latent correlations. We treat the correlations computed by the ORG approach as gold
standard and evaluate the maximum value of the absolute difference with the latent correlation estimates using the two approximation schemes, ML
and MLBD.

\subsubsection{Comparison on simulated data}
To assess the approximation accuracy in simulations, we generate two variables using five types combinations: TC, TT, BC, BB, and TB. Here we present results for the TC case, other cases are available in the Supplementary Material.  First, we generate two Gaussian variables of sample size $n=100$ with mean $0$ and fixed value of latent correlation (we consider nine values from 0.05 to 0.91). Given the zero proportion value $\pi_0$ (we consider eleven values from 0.03 to 0.95), we shift both variables so that the truncation applied at zero leads to desired value of $\pi_0$. That is, we truncate one of the variables by zeroing all negative values that remain after the shift. 

\begin{figure}[!t]
    \centering
    \includegraphics[scale = 0.7]{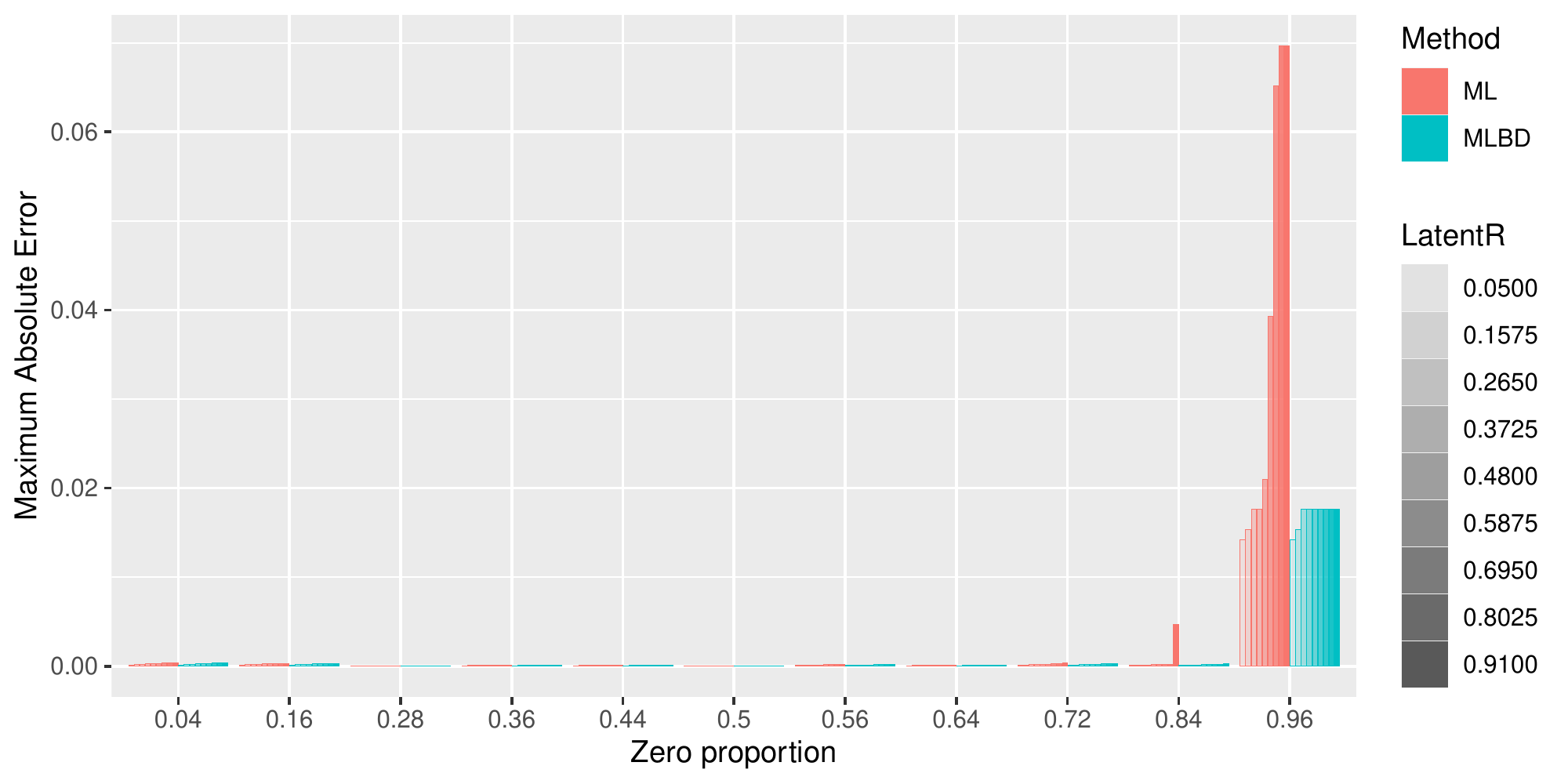}
    \caption{TC case. Maximum absolute error of multilinear interpolation approach (ML) and hybrid estimation approach (MLBD) for two simulated variables of sample size $n=100$ across 100 replications. One variable is truncated continuous with zero proportion levels from 0.04 to 0.96 (T) and the other variable is continuous (C). 
    }\label{fig:accuracy_TC}
\end{figure}

Figure~\ref{fig:accuracy_TC} shows the maximum absolute error between the approximated values 
using interpolation and the gold standard values estimated by optimizing bridge inverse 
function across 100 replications. In Figure~\ref{fig:accuracy_TC}, the highest maximum
absolute error for the ML method is $0.0406$ at latent $r=0.91$ and zero proportion rate 
$\pi_0=0.95$, respectively. The MLBD method reduced the error to $0.0101$. When $\pi_0=0.858$, 
ML's maximum error is only $0.0022$. All other maximum absolute errors are less than or equal 
to $0.0004$ and, on average, $0.0002$, thus suitable for downstream statistical inference. 

In summary, the approximation error for the TC case increases with the 
increase in zero proportion. However, the MLBD method accounts for the extreme cases, 
leading to significantly smaller approximation error compared to ML. The results for TT, 
BC, BB, and TB cases are similar (Supplementary Material). The approximation error 
increases with the increase in proportion of zeros for the truncated variable. The 
approximation error also increases as the binary variable gets more unbalanced in the 
number of zeros and ones. In all cases, the approximation error for the MLBD method is 
better compared to ML method.

\subsubsection{Comparison on real data}
\label{sec:acc_real}
We next consider three real-world data sets. The first data set is a subset of the quantitative 
microbiome profiling data (QMP), put forward in \citet{Vandeputte2017}, which comprises $n=106$ 
samples across $p=91$ bacterial genera, resulting in a $91$ by $91$
latent correlation matrix estimation problem. The second data set is taken 
from the American gut project (AGP) \citep{McDonald2018} and comprises filtered amplicon data for $p=481$
operational taxonomic units (OTUs) across $n=6482$ samples. Both microbiome data sets are treated as
truncated continuous, and we use the bridge inverse function for the TT case. 
The final dataset is based on multi-view data from TCGA-BRCA (the cancer genome atlas breast invasive
carcinoma) project, comprising gene expression data of 891 genes and micro RNA data
of 431 micro-RNAs across 500 samples. The gene expression data are treated as continuous, and the
micro-RNA data as truncated continuous. The latent correlation matrix for the gene expression data (of
size 891 by 891) can be calculated using the explicit form of the bridge inverse function for the CC
case. The correlation matrix estimates between micro-RNAs and genes (of size 431 by 891) and
between micro-RNAs (of size 431 by 431) are calculated using the TC and the TT
bridge inverse functions, respectively. The entire latent correlation estimate is of size 1322 by 1322
($891+431=1322$). 

We observed that, in the QMP and AGP data, there are no pairs of variables outside of our boundary
specification~\eqref{eq:boundary}, implying that ML and MLBD give identical estimates. In the micro-RNA data in TCGA-BRCA,
six pairs of variables are outside of the specified bounds. We observed  that maximum absolute error
between ML (and MLBD) to the gold standard is $0.0006$ on both the QMP data
and the AGP data. The maximum error for MLBD on the TCGA-BRCA data is $0.0005$. MLBD's
mean absolute error is 8.0e-05, 7.3e-05, and 1.1e-05 on QMP, AGP, and TCGA-BRCA data, respectively.

\subsubsection{Comparison for graphical model estimation}\label{sec:acc_spring}
We next assess the MLBD scheme in the context of sparse graphical model estimation with 
SPRING (Semi-Parametric Rank-based approach
for INference in Graphical model) \citep{SPRING}. SPRING uses latent correlation estimation followed by 
neighborhood selection \citep{MBmethod} to estimate sparse graphical models from quantitative and relative
microbial abundance data. SPRING selects the optimal tuning parameter $\lambda$ level via the Stability Approach
to Regularization Selection (StARS) \citep{Liu2010} which requires
repeated subsampling of the data to estimate edge selection probabilities and thus repeated latent correlation matrix estimation. 

To assess MLBD's approximation accuracy, we measured the absolute difference of the entries in the
estimated sparse partial correlation matrices between ORG and MLBD across two different regularization paths ($\lambda$-paths). We set the number of subsamples to 50. 
We first considered a fixed
$\lambda$-path with 50 values log-linearly spaced in the interval $[0.006,0.6]$ for both schemes. At the StARS-selected
$\lambda_\text{StARS}$, we observed a maximum absolute difference of $0.0010$ and mean difference of 7.4e-06, respectively, in the
resulting partial correlation estimates. We also used a data-driven regularization path comprising 50 $\lambda$ values,
log-linearly spaced in $[0.01\sigma_{\max},\sigma_{\max}]$ where $\sigma_{\max}$ is the largest off-diagonal element in the respective
latent correlation estimates ($\sigma_{\max}=0.8183$ for ORG, and $\sigma_{\max}=0.8186$ for MLBD, respecitly). At the StARS-selected
$\lambda_\text{StARS}$ value, we observed a maximum difference of $0.0011$ and a mean error of 7.8e-06 in the resulting partial correlation estimates.


\subsection{Computational Speed-up}
We report the numerical run times and highlight the speed-up of our approximation scheme 
on all described test scenarios. Run times are measured using the \textsf{microbenchmark}
\textsf{R} package on a 
Linux system with Intel(R) Xeon(R) CPU E5-2680 v4 @ 2.40GHz.
Table~\ref{tab:twovar_comptime} presents run time results (in microseconds ($\mu s$)) for
the synthetic data scenarios. Here, we consider pairs of simulated variables for all five
data type combinations. 

\begin{table}[t]
\caption{Run time (in microseconds $[\mu s]$) for latent correlation estimation across variable pairs (C - continuous, B - binary, T - truncated). \label{tab:twovar_comptime}}

			\centering
			\begin{tabular}{rrrrrr}
				\hline
				& TC & TT & BC & BB & TB \\ 
				\hline
				ORG & 3767.28 & 24255.43 & 2516.40 & 2894.14 & 2177.13 \\ 
				ML & 350.72 & 454.88 & 352.73 & 446.21 & 452.80 \\ 
				MLBD & 362.72 & 479.92 & 368.92 & 483.67 & 493.77 \\ 
				\hline
			\end{tabular}

\end{table}

The run time of the ORG method is highly data type dependent. For instance, the TT case,
which is relevant for amplicon, Chip-Seq, or single-cell data, has the longest run time
($\sim 24255 \mu s$) due to the four-dimensional normal cdfs in its bridge function. Here,
the ML and MLBD methods achieve a speed-up of about $50x$. For the other cases, both
approximation schemes achieve a $4x - 10x$ speed-up compared to the direct optimization
scheme. As expected, the run time of the hybrid MLBD scheme is longer than ML but allows
tight control of approximation errors when estimated Kendall's $\tau$ values fall outside the boundary (see~\eqref{eq:boundary}). 

\begin{table}[t]
\caption{Run time (in seconds [s]) for latent correlation estimation on biological data. \label{tab:comptime_combined}}
\centering
			\begin{tabular}{ccccc}
			\hline 
	& \multicolumn{3}{c}{latent correlation} & \multirow{2}{*}{SPRING on QMP}\\
			\cline{2-4}
			& QMP & AGP & TCGA-BRCA & \\
		\hline 
ORG & 59.63$^\dagger$ & 3459.05$^\S$ & 2039.52$^\S$ & 1810.39$^\S$ \\ 
  MLBD & 0.97$^*$ & 320.56$^\dagger$ & 245.13$^\dagger$ & 68.06$^\S$ \\ 
  Kendall & 0.94$^*$ & 318.09$^\dagger$ & 200.27$^\dagger$ &   \\ 
		\hline 
	\end{tabular}\\
	$^*$: median value over 100 repetitions and $^\dagger$: median value over 10 repetitions, and $^\S$: one time result. 
\end{table}

Table~\ref{tab:comptime_combined} shows the run time results for latent correlation and
graphical model estimation. For comparison, we also include run time results for computing
Kendall's $\tau$ matrix using \textsf{cor.fk} function in \textsf{R} package \textsf{pcaPP}. We observe that
MLBD achieves significant speed up of between $8x$ for the TCGA-BRCA data to more than $60x$ on the QMP
data. In addition, MLBD's computational cost is comparable to plain Kendall's $\tau$ calculation for the AGP
and QMP data, and is only $1.2x$ slower on the TCGA-BRCA data.

\begin{figure}[t]
    \centering
    \includegraphics[width = 0.6\textwidth]{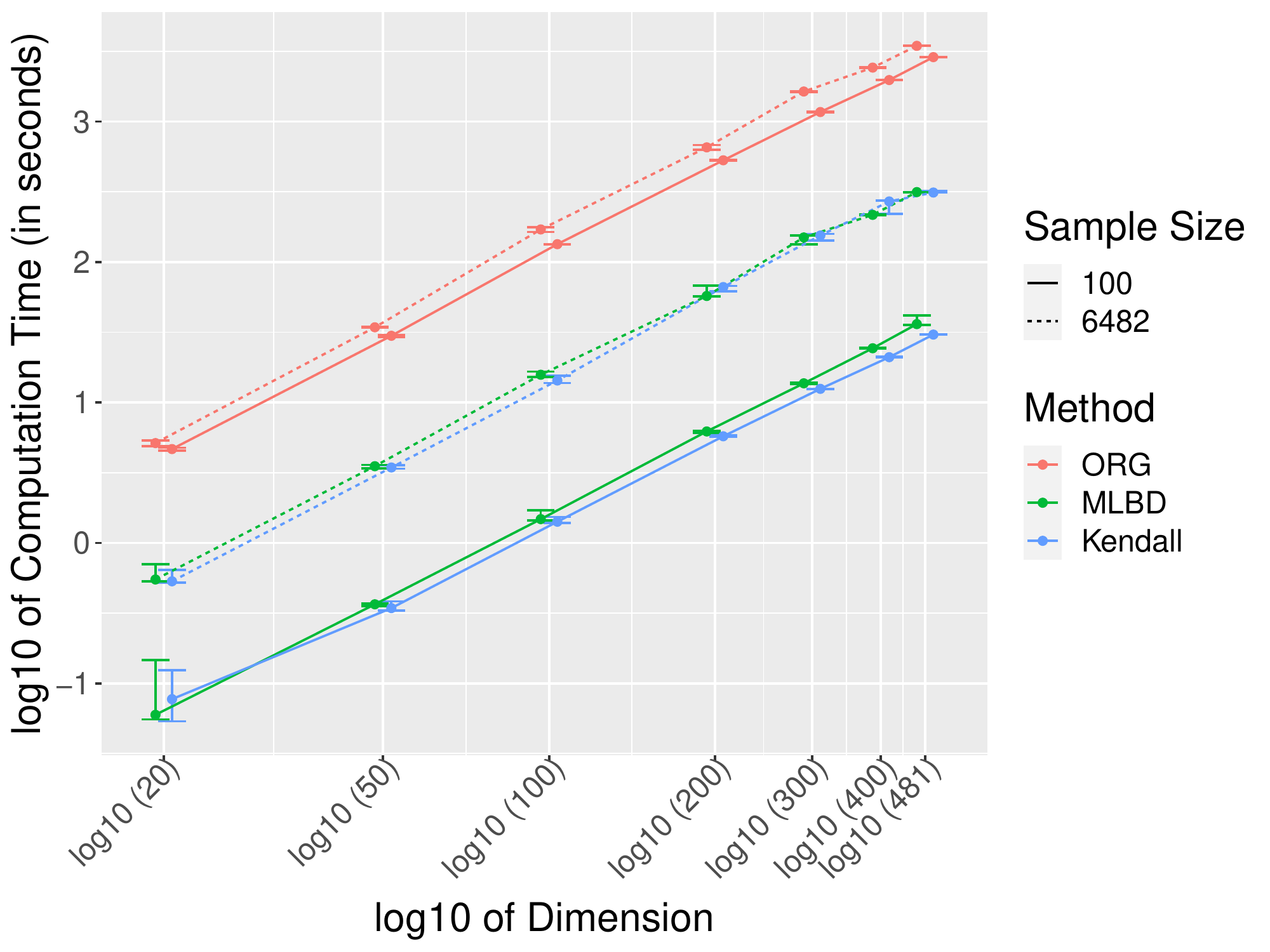}
    \caption{Computational scaling of the run time (median and standard deviation in log10 scale, in seconds) versus dimension $p$ (in log10 scale) for the original optimization method (ORG, two repetitions), the proposed hybrid multi-linear interpolation method (MLBD, TT case, ten repetitions), and Kendall's $\tau$  (Kendall, ten repetitions). The Amplicon data from AGP is used for two different sample sizes, $n=100$ (solid) and $n=6482$ (dotted). All methods show the expected $O(p^2)$ complexity as reflected in the linear run time increase with slope $\approx 2$ in the log-log plot. MLBD is one order of magnitude faster than ORG and comparable in run time to standard Kendall's $\tau$.}
    \label{fig:runtime}
\end{figure}

We next investigate the run time scaling behavior of the ORG, MLBD (using the TT case), and Kendall's
estimators with increasing dimensions $p = [20, 50, 100, 200, 300, 400]$ at two different sample sizes
$n=100, 6482$ using the AGP data. Figure~\ref{fig:runtime} summarizes the observed scaling in a log-log plot. 
For all methods we observe the expected $O(p^2)$ scaling behavior with dimension $p$, i.e., a linear scaling in the log-log
plot. However, MLBD is at least one order of magnitude faster than ORG and comparable in run time to standard Kendall's $\tau$ independent of the dimension of the problem.

\section{Discussion}
\label{sec:conc}

We have introduced a fast method for computing latent correlations for variable pairs of continuous/binary/truncated types.
The method is implemented in the R package \textsf{mixedCCA} and allows the estimation of latent correlations at a
computational cost that is similar to standard Kendall's $\tau$ computation. Several improvements of the method can be pursued. First, the hybrid MLBD method uses a boundary condition $|\widehat \tau| > 0.9\text{ABD}$, however the constant 0.9 can be adapted for specific variable combination (TT, TC, TB, BB or TT). Our simulation studies (Supplementary Material) suggest that a stricter boundary (lower constant) may be needed for the BC case to achieve similar approximation error as for the TC case. Another alternative is to use a boundary based on the zero proportion value only. However, this is a conservative approach as it ignores the dependence of approximation error on the size of latent correlation. For example, if the latent correlation $r = F^{-1}(\tau, \Delta)$ is very close to 0, the bound in Theorem~\ref{thm:BCbound} is not as large for the same value of zero proportion as when $|r|$ is close to one. Secondly, motivated by the need for fast and accurate methods for processing modern high-dimensional sequencing data, we have 
focused here on processing sparse, highly skewed count or binary data. For ordinal variable types \citep{Quan:2018tv,
Feng:2019wh}, which also have non-trivial bridge functions, our interpolation approach will likely also achieve faster
latent correlation computation. Finally, in its current form, our hybrid multilinear interpolation scheme requires storing
pre-computed function values on a large grid of points. An alternative potentially fruitful approach is to construct a
closed-form analytical function that approximates the inverse bridge function directly, thus completely eliminating the
grid. The shape of the inverse bridge function for the TC case (Figure~\ref{fig:bridgeTCinv_3d}) suggests that sigmoid
log-logistic approximation functions \citep[Chapter~3]{kyurkchiev2015sigmoid} could be promising candidates, since they can
adapt their smoothness to mimic the observed change from the sinusoidal function (zero proportion is equal to zero in
Figure~\ref{fig:bridgeTCinv_3d}) to the step function (zero proportion is equal to one). We leave these investigations for
future research.

\bigskip
\begin{center}
{\large\bf SUPPLEMENTARY MATERIAL}
\end{center}

\begin{description}

\item[Supplementary:] Proofs of Theorem~\ref{thm:BCbound} and~\ref{thm:TCbound}, derivation of bound~\eqref{eq:boundary}, description of interpolation grid and additional approximation accuracy results for the TT, TB, BC and BB cases (.pdf file)


\end{description}


\bibliographystyle{biomAbhra}

\bibliography{Bfast}
\end{document}


\def\spacingset#1{\renewcommand{\baselinestretch}%
{#1}\small\normalsize} \spacingset{1}


\if0\blind
{
  \title{\bf Supplementary material for ``Fast computation of latent correlations''
}
  \author{Grace Yoon, Christian L. M\"{u}ller and Irina Gaynanova\\
    }
  \maketitle
} \fi

\spacingset{1.5} 


\section{Description of interpolation grid}

In our numerical implementation, we use the same grid for both Algorithms~\ref{alg:ML} and~\ref{alg:MLBD}, available in the R package~\textsf{mixedCCA} (\url{https://github.com/irinagain/mixedCCA}). For $\tau$, we use  $\textsf{seq}(-0.99, 0.99, \textsf{by = } 0.01)$ for the TC and TT cases. For the BC, BB and TB cases, we construct the positive part of the grid as $\tau_1 = \textsf{c}(\textsf{seq}(0.001, 0.095, \textsf{by = } 0.005))$, $\textsf{seq}(0.101, 0.5, \textsf{by = } 0.007)$, and then we use $\textsf{c}(-\textsf{rev}(\tau_1), 0, \tau_1)$. The grid for $\tau$ in BC, BB and TT cases is thus more dense around zero values as those cases have more zero pairs, leading to the decrease in the range of $\tau$. For $\Delta$, we first construct the grid based on the values of $\pi_0$ and then convert to $\Delta$
using the inverse of normal cdf: $\Phi^{-1}\left(\log10\left\{ \textsf{seq}\left(1, 10^{0.99}, \textsf{length.out = }
50\right)\right\}\right)$ for truncated type and $\Phi^{-1}\left( \textsf{seq}\left(0.01, 0.99, \textsf{length.out = } 50\right)\right)$
for the binary type, respectively. The grid for truncated type is skewed to have denser grid around higher values of $\pi_0$ and coarser grid around lower value of $\pi_0$ as the interpolation error decreases as the proportion of zeros decreases. The grid sizes were chosen to satisfy 5 MB restriction on the size of R packages on CRAN.

\section{Approximation accuracy in calculation of latent correlation in simulation}\label{sec:addaccfig}

\begin{figure}[!t]
    \centering
    \includegraphics[scale = 0.7]{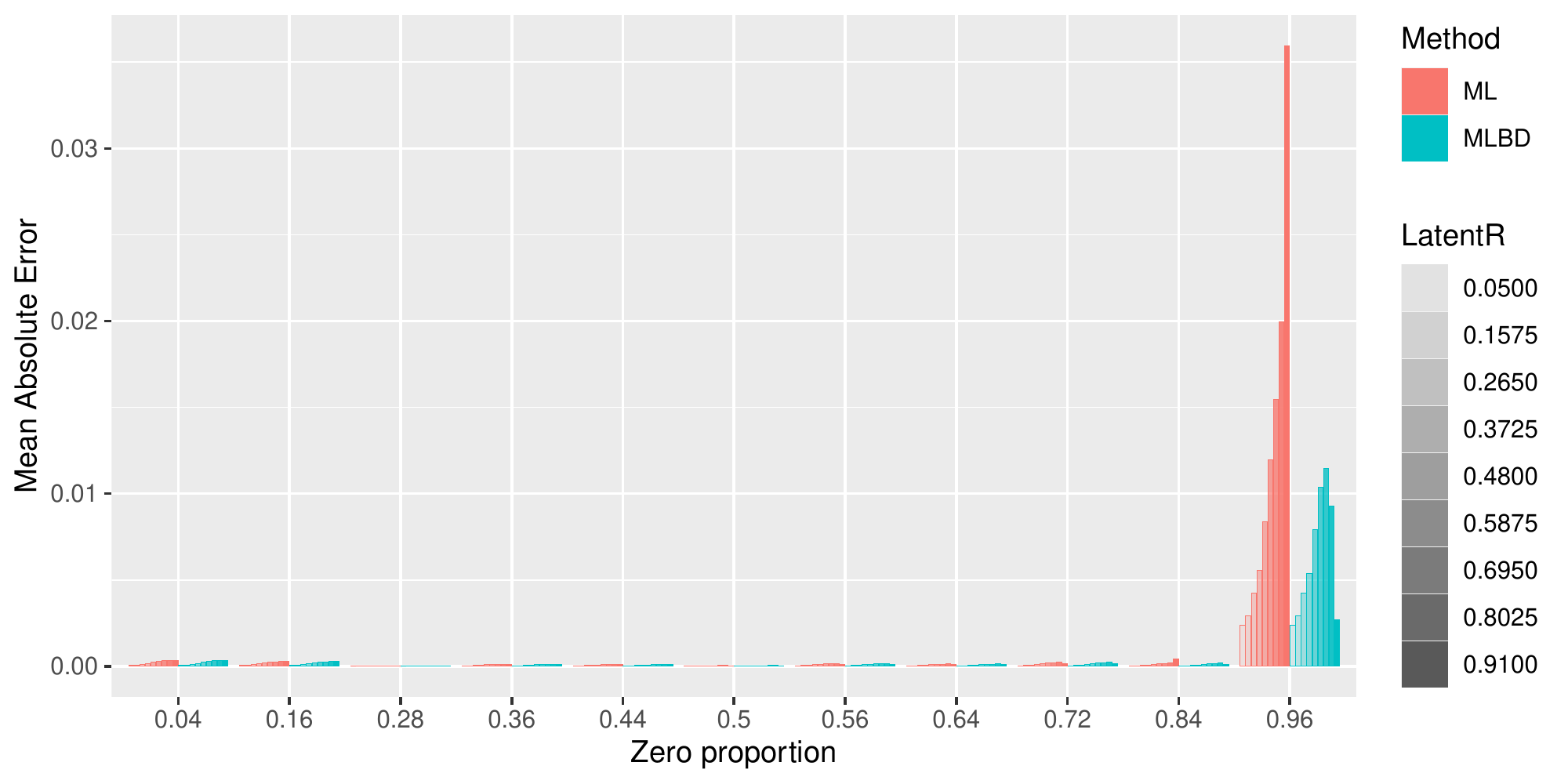}
    \caption{TC case. Mean absolute error of multilinear interpolation approach (ML) and hybrid estimation approach (MLBD) as described in Section~\ref{sec:hybridcomp} for two simulated variables of sample size $n=100$. The y-axis represents the mean absolute error across 100 replications. One variable is truncated continuous type with varied zero proportion levels from 0.04 to 0.96 shown on x-axis and the other variable is continuous type.}
    \label{fig:accuracy_TC_meanAD}
\end{figure}

\begin{figure}[!t]
    \centering
    \includegraphics[scale = 0.7]{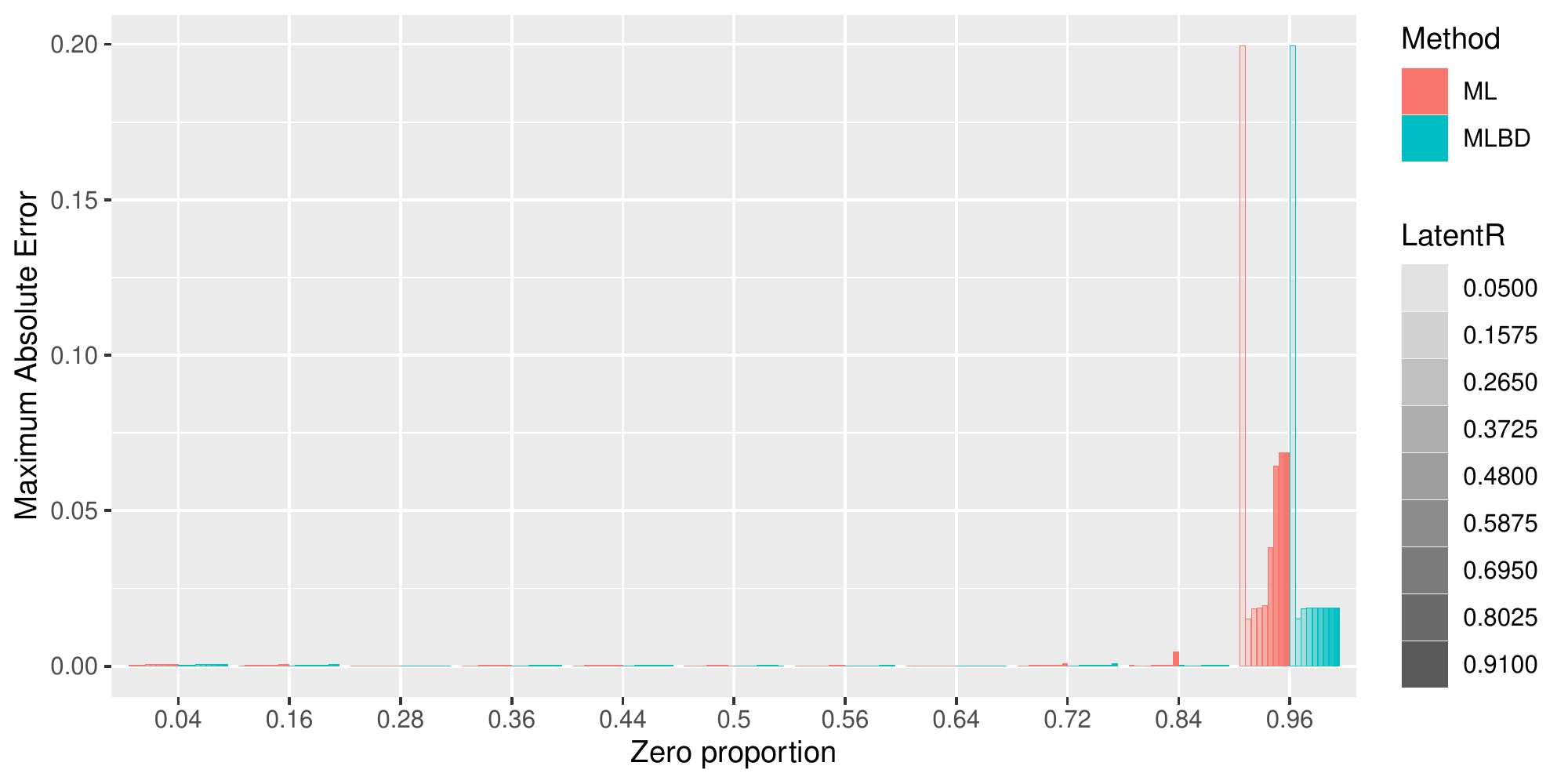}
    \includegraphics[scale = 0.7]{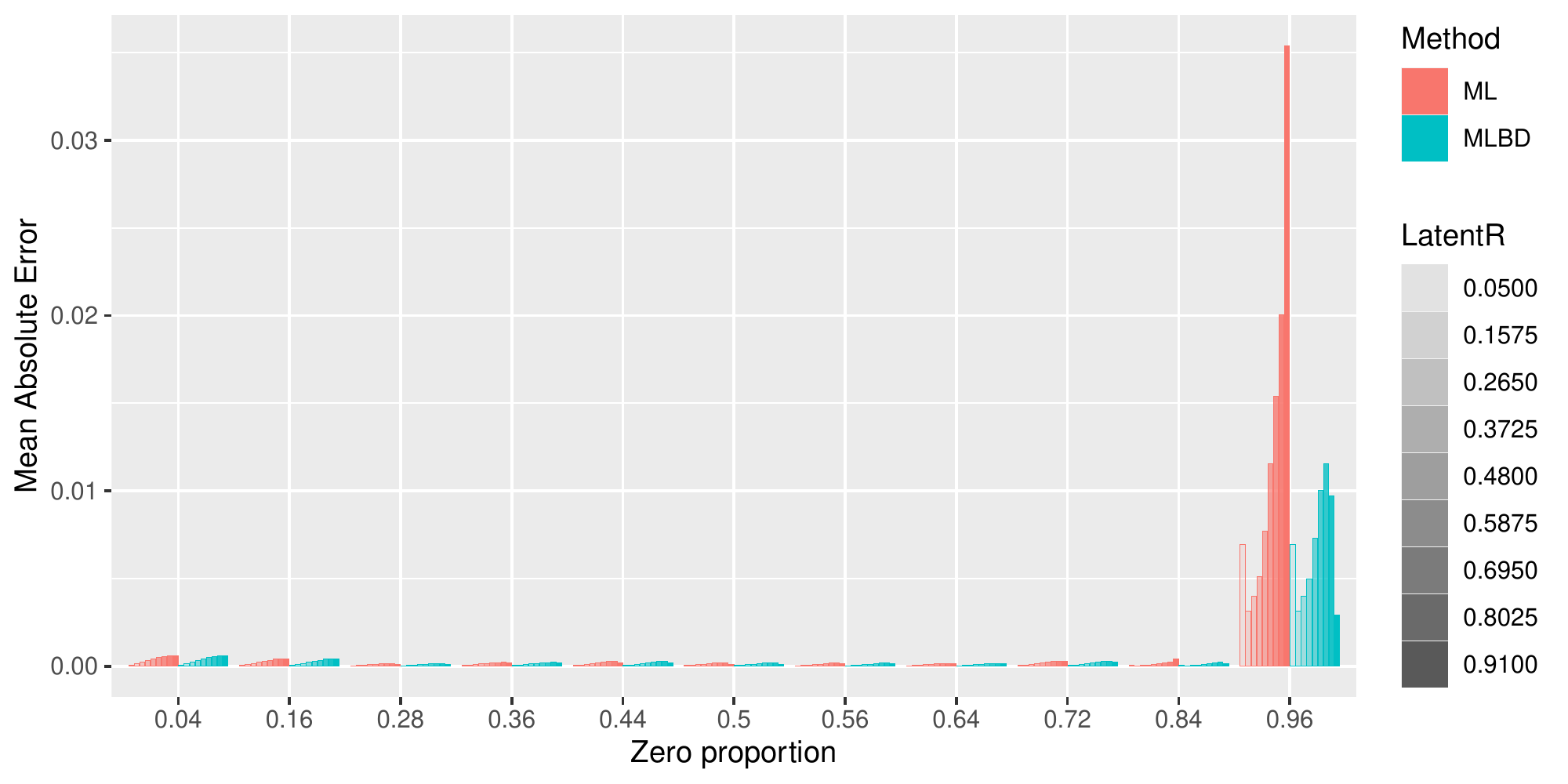}
    \caption{TT case. Maximum absolute error \textbf{(Top)} and mean absolute error \textbf{(Bottom)} of multilinear interpolation approach (ML) and hybrid estimation approach (MLBD) as described in Section~\ref{sec:hybridcomp} for two simulated variables of sample size $n=100$. The y-axis represents the mean absolute error across 100 replications. One variable is truncated continuous type with zero proportion $\pi_0$ and the other variable is truncated continuous type with zero proportion $\pi_0/2$, where $\pi_0$ is changing from 0.04 to 0.96 shown on x-axis.}
    \label{fig:accuracy_TT}
\end{figure}

\begin{figure}[!t]
    \centering
    \includegraphics[scale = 0.7]{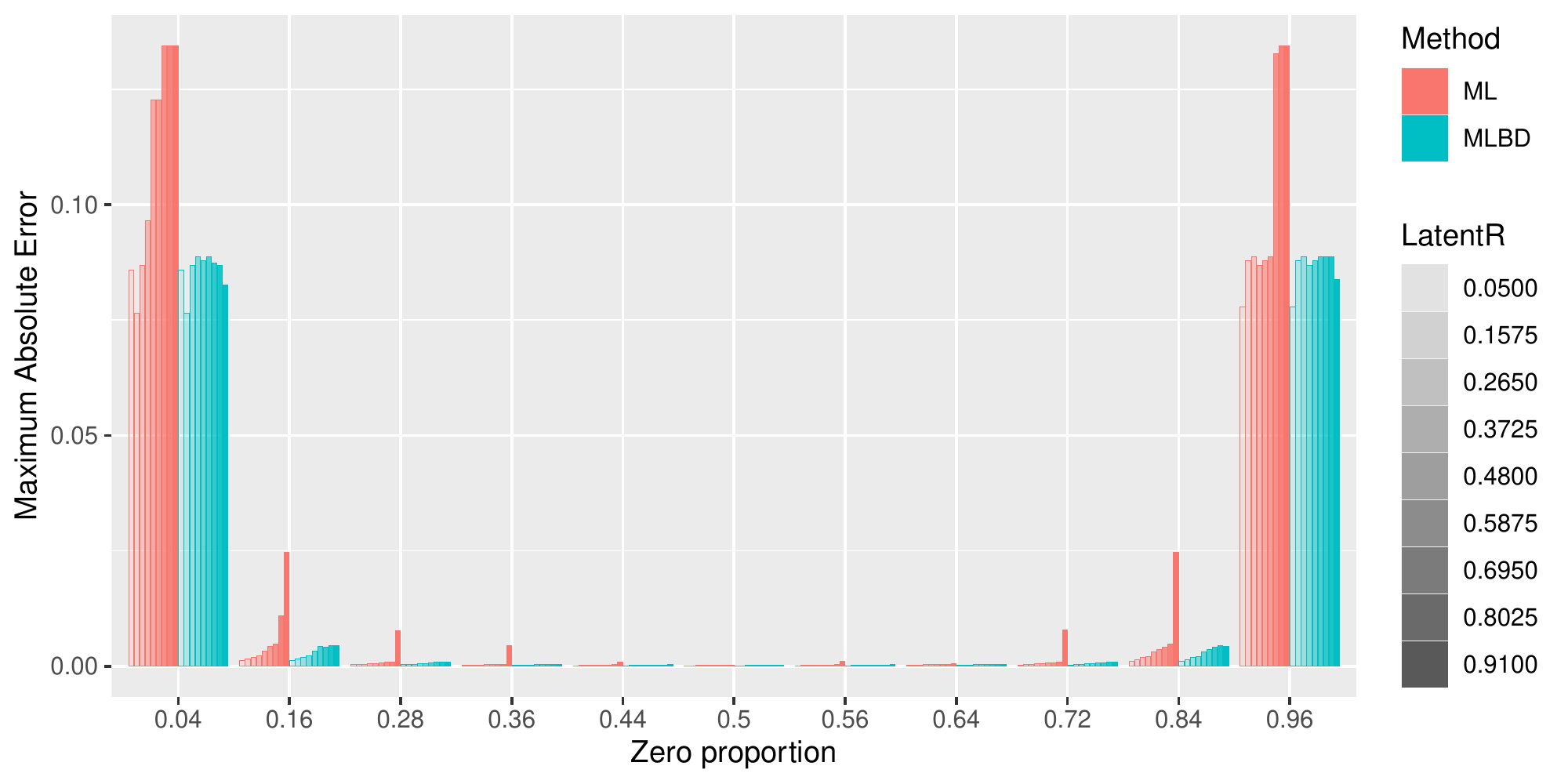}
    \includegraphics[scale = 0.7]{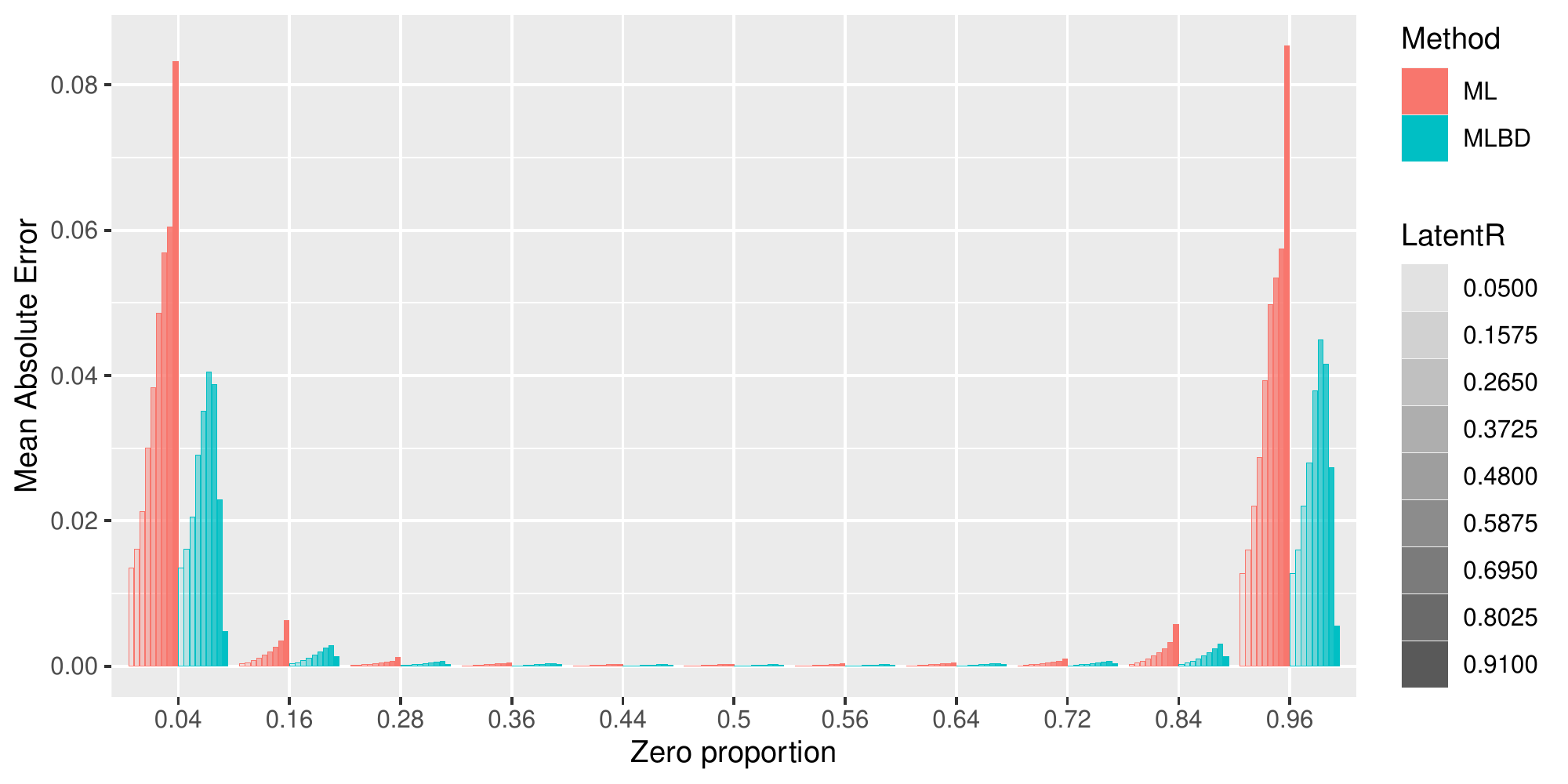}
    \caption{{BC case.} Maximum absolute error \textbf{(Top)} and mean absolute error \textbf{(Bottom)} of multilinear interpolation approach (ML) and hybrid estimation approach (MLBD) as described in Section~\ref{sec:hybridcomp} for two simulated variables of sample size $n=100$. The y-axis represents the mean absolute error across 100 replications. One variable is binary type with varied zero proportion levels (x-axis) and the other variable is continuous type.}
    \label{fig:accuracy_BC}
\end{figure}

\begin{figure}[!t]
    \centering
    \includegraphics[scale = 0.7]{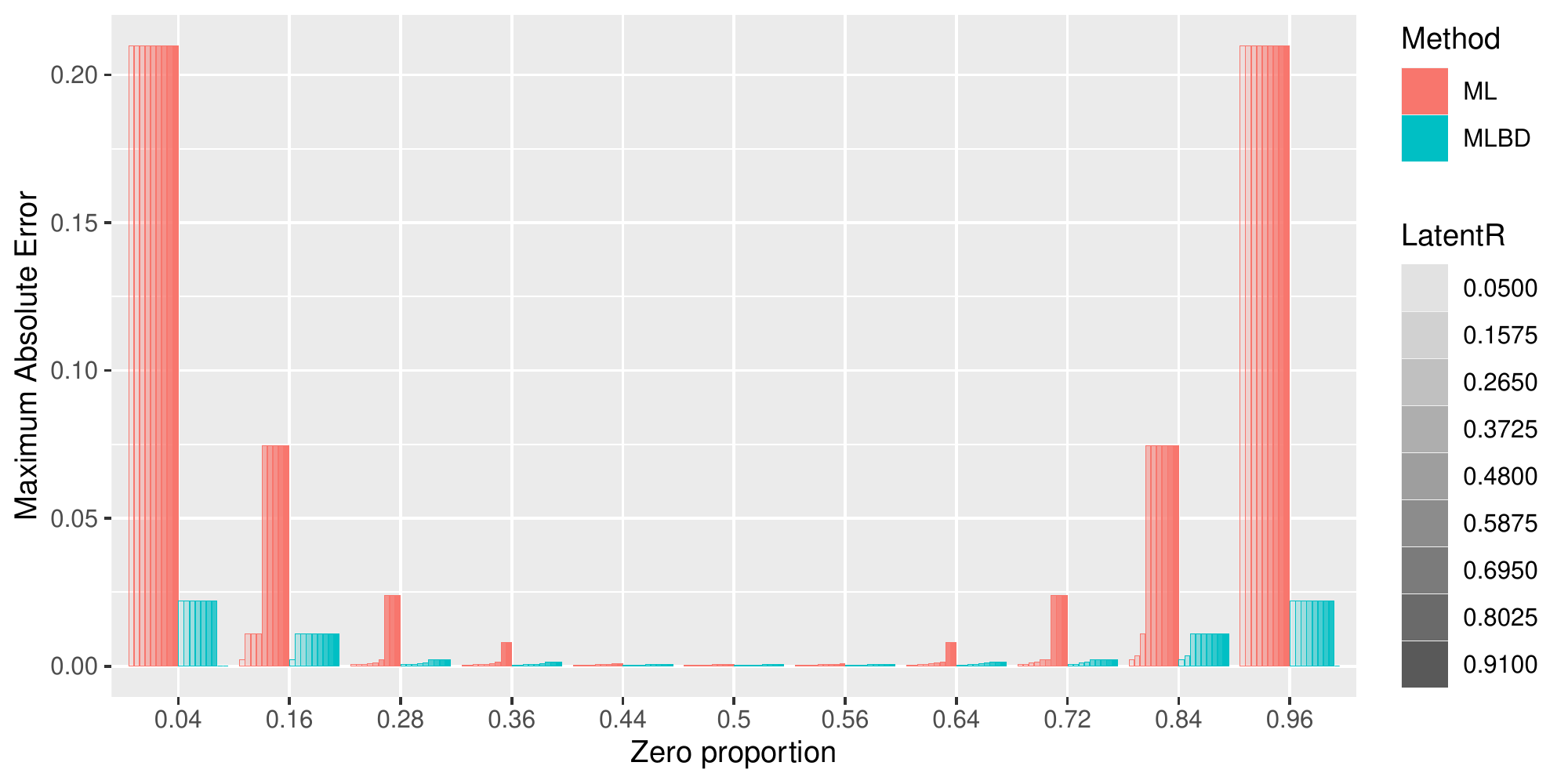}
    \includegraphics[scale = 0.7]{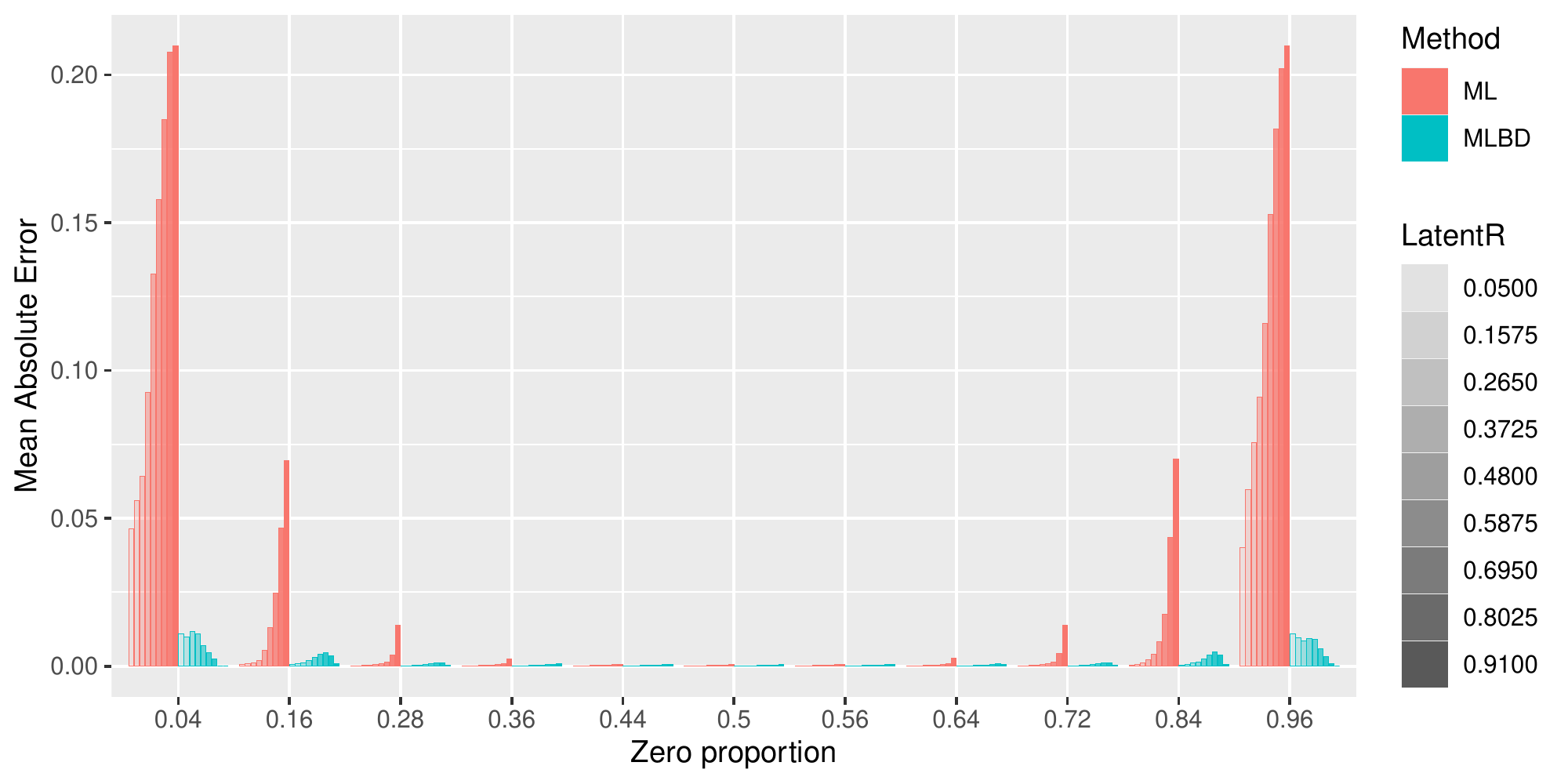}
    \caption{BB case. Maximum absolute error \textbf{(Top)} and mean absolute error \textbf{(Bottom)} of multilinear interpolation approach (ML) and hybrid estimation approach (MLBD) as described in Section~\ref{sec:hybridcomp} for two simulated variables of sample size $n=100$. The y-axis represents the mean absolute error across 100 replications. One variable is binary type with varied zero proportion levels (x-axis) and the other variable is also binary type with 0.5 fixed zero proportion level.}
    \label{fig:accuracy_BB}
\end{figure}

\begin{figure}[!t]
    \centering
    \includegraphics[scale = 0.7]{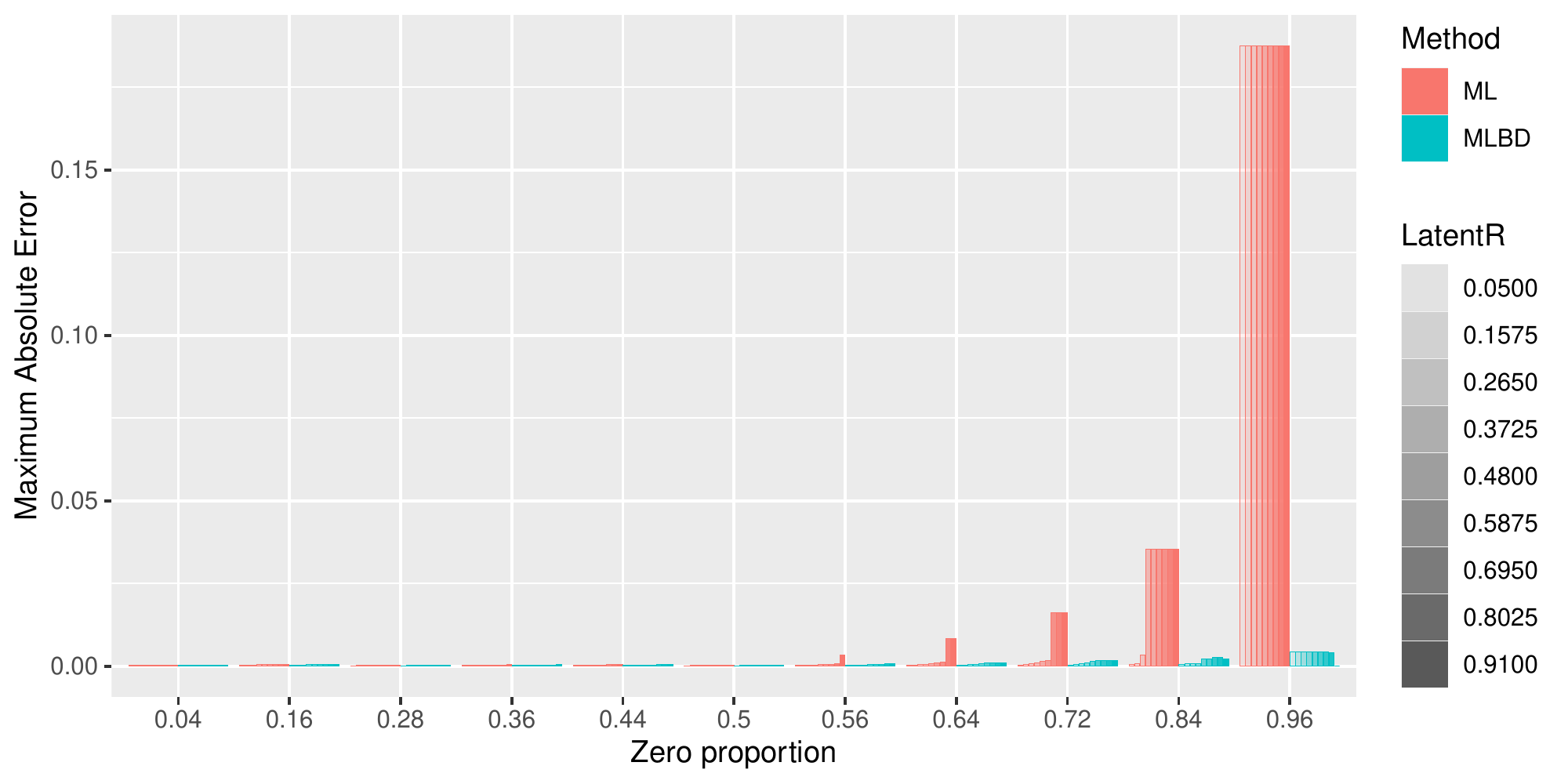}
    \includegraphics[scale = 0.7]{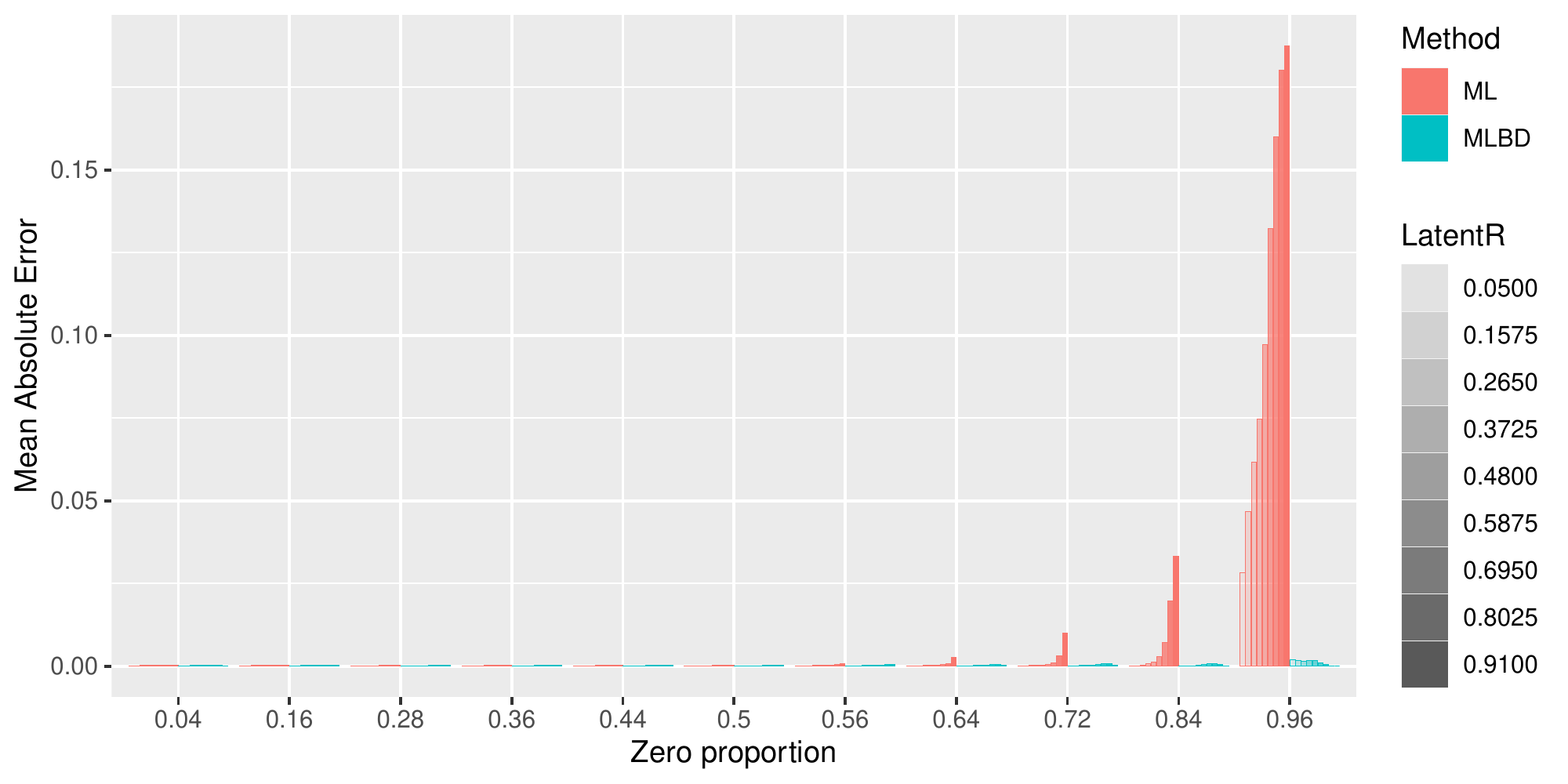}
    \caption{TB case. Maximum absolute error \textbf{(Top)} and mean absolute error \textbf{(Bottom)} of multilinear interpolation approach (ML) and hybrid estimation approach (MLBD) as described in Section~\ref{sec:hybridcomp} for two simulated variables of sample size $n=100$. The y-axis represents the mean absolute error across 100 replications. One variable is truncated continuous type with varied zero proportion levels (x-axis) and the other variable is binary type with 0.5 fixed zero proportion level.}
    \label{fig:accuracy_TB}
\end{figure}

In this section, we complement the results of Section~\ref{sec:perfo} with TC, TT, BC, BB and TB cases using the same data generation mechanism. 
Figure~\ref{fig:accuracy_TC_meanAD} shows the mean absolute error of two approximation methods, ML and MLBD, in the TC cases. Figures~\ref{fig:accuracy_TT}-\ref{fig:accuracy_TB} shows maximum and mean absolute error for TT, BC, BB and TB cases. For binary variables, we find a quantile of the variable based on the specified zero proportion value, and then dichotomize the data by setting the value to one if it is larger than the quantile and zero otherwise. As expected, the approximation error increases with the increase in proportion of zeros for truncated variable. The approximation error also increases as the binary variable gets more in-balance in the number of zeros and ones. The approximation error for MLBD method is always better than the approximation error for ML method.  A reproducible workflow of the presented numerical results is 
available at \url{https://github.com/GraceYoon/Fast-latent-correlation}.

\clearpage

\section{Derivation of boundary region for multilinear interpolation}\label{sec:bound}

The sample Kendall's $\tau$ formula \eqref{eq:tau} in the main manuscript compares signs of all possible pairs $n(n-1)/2$ for sample size $n$. By subtracting how many zero pairs occur from total number of pairs, we approximate ranges of Kendall's $\tau$ for the TT, BC, BB and TB cases.

\subsection{TT case}
Let $\bx\in \R^n$ and $\by\in \R^n$ be the observed $n$ realizations of two truncated continuous type variables. Let $n_{0x} = \sum_{i=1}^{n} I(x_{i}=0)$, $n_{0y} = \sum_{i=1}^{n} I(y_{i}=0)$ be the number of zeros in each variable, and $n_{0\text{both}} = \sum_{i=1}^{n} I(x_{i}=0 ~\&~ y_{i} = 0)$ be the number of samples having concurrent zeros in both variables. We first find the upper bound by subtracting how many pairs are possible between zeros in each variable from the total number of possible pairs and adding back the number of pairs between zeros where both variables are zeros based on the general addition rule in set operations. 
	\begin{equation}
	    \begin{split}\nonumber
	        | \tau(\bx, \by) | & \le \dfrac{\bpm n \\ 2 \epm - \bpm n_{0x} \\ 2 \epm - \bpm n_{0y} \\ 2 \epm + \bpm n_{0{\rm both}} \\ 2 \epm }{\bpm n \\ 2 \epm}
	    \end{split}
	\end{equation}
Since $n_{0\text{both}} \le \min(n_{0x}, n_{0y})$, we obtain
	\begin{equation}
	    \begin{split}\nonumber
	        | \tau(\bx, \by) |
	        & \le \dfrac{\bpm n \\ 2 \epm - \bpm n_{0x} \\ 2 \epm - \bpm n_{0y} \\ 2 \epm + \bpm \min(n_{0x}, n_{0y}) \\ 2 \epm }{\bpm n \\ 2 \epm} \le \dfrac{\bpm n \\ 2 \epm - \bpm \max(n_{0x}, n_{0y}) \\ 2 \epm }{\bpm n \\ 2 \epm}\\
	        & \le 1 - \dfrac{ \max(n_{0x}, n_{0y}) ( \max(n_{0x}, n_{0y}) - 1)}{n(n - 1)}\\
	        & \approx 1 - \left(\dfrac{ \max(n_{0x}, n_{0y }) }{n}\right)^2 = 1 - \left\{ \max(\pi_{0x}, \pi_{0y}) \right\}^2 
	    \end{split}
	\end{equation}
where $\pi_{0x} = n_{0x}/n$ and $\pi_{0y} = n_{0y}/n$.

\subsection{BC case}
Let $\bx\in \R^n$ and $\by\in \R^n$ be the observed $n$ realizations of binary and continuous variables, respectively. Let $n_{0} = \sum_{i=1}^{n} I(x_{i}=0)$ and $\pi_{0} = n_{0}/n$. In this case, we exclude the ties, and find the upper bound by only counting the pairs having one value of the pair as zero and the other value as one. That is,
	\begin{equation}
	    \begin{split}\label{eq:up-bc}
	        	| \tau(X,Y) | & \le \left\{ n_{0} (n-n_0) \right\} \Big/ \bpm n \\ 2 \epm  = 2\dfrac{n_0 (n - n_0)}{n(n-1)} = 2\left(\dfrac{n_0}{n}\right) \left( \dfrac{n-n_0}{n-1} \right) \\
	        	& \approx  2\left(\dfrac{n_0}{n}\right) \left( 1- \dfrac{n_0}{n} \right) = 2\left(\pi_0\right) \left( 1- \pi_0 \right).
	    \end{split}
	\end{equation}

\subsection{BB case}
For binary variable, we know $\pi_{0} = 1 - \pi_1$, where $\pi_1$ is the proportion of zero values and $\pi_1$ is the proportion of one values. Rewriting \ref{eq:up-bc} as $2\left(1-\pi_1\right) \left( 1- \pi_0 \right)$, we find the upper bound for BB case by taking the maximal proportion between the two variables:
	\begin{equation}
	    \begin{split}\nonumber
	        	| \tau(\bx,\by) | & \le 2\left(1- \max(\pi_{1x}, \pi_{1y}) \right) \left( 1- \max(\pi_{0x}, \pi_{0y}) \right).
	    \end{split}
	\end{equation}

\subsection{TB case}

Let $\bx\in \R^n$ and $\by\in \R^n$ be the observed $n$ realizations of truncated and binary variables, respectively. Let $n_{0x} = \sum_{i=1}^{n} I(x_{i}=0)$, $\pi_{0x} = n_{0x}/n$, $n_{0y} = \sum_{i=1}^{n} I(y_{i}=0)$, $\pi_{0y} = n_{0y}/n$ and  $a = \sum_{i=1}^{n} I(x_i = 0 ~\&~ y_i = 0)$. Then, the number of nonzero pairs is 
\begin{equation}
    \begin{split}\nonumber
        n_{0y} ( n - n_{0y}) - a(n_{0x} - a) & = n n_{0y} - n_{0y}^2 - a n_{0x} + a^2\\
        & = \left( a - \dfrac{n_{0x}}{2} \right)^2 -  \dfrac{n_{0x}^2}{4} + n_{0y}(n - n_{0y}),
    \end{split}
\end{equation}
and this number of nonzero pairs is maximized at one of the extreme values of $a$. 

First, if $n_{0x} > \max \left( n_{0y}, n - n_{0y} \right)$, $a$ satisfies $ n_{0x} - n +n_{0y} \le a \le n_{0y}$. With $a = n_{0x} - n +n_{0y}$, 
\begin{equation}
    \begin{split}\nonumber
        n_{0y} ( n - n_{0y}) - a(n_{0x} - a) & \le  n_{0y} ( n - n_{0y}) - (n_{0x} - n +n_{0y})(n - n_{0y})\\
        & = (n-n_{0x})(n - n_{0y}).
    \end{split}
\end{equation}
With $a = n_{0y}$,
\begin{equation}
    \begin{split}\nonumber
        (n_{0y}) ( n - n_{0y}) - a(n_{0x} - a) & \le  n n_{0y} - n_{0y}^2 - n_{0y} n_{0x} + n_{0y}^2
         =  n_{0y} (n -  n_{0x}).
    \end{split}
\end{equation}
Thus, the number of nonzero pairs is bounded by
\begin{equation}
    \begin{split}\nonumber
        n_{0y} ( n - n_{0y}) - a(n_{0x} - a) & \le \max(n_{0y}, n - n_{0y}) \times (n - n_{0x}).
    \end{split}
\end{equation}
Second, if $n_{0x} \le \max \left( n_{0y}, n - n_{0y} \right)$, either $a = n_{0x}$ or $a=0$ maximizes the number of nonzero pairs, which leads to the same upper bound $n_{0y} (n -  n_{0y})$ as the BC case.

Combining the first and the second cases, the total number of nonzero pairs is bounded by
\begin{equation}
    \begin{split}\nonumber
        n_{0y} ( n - n_{0y}) - a(n_{0x} - a) & \le \max(n_{0y}, n - n_{0y}) \times (n - \max(n_{0y}, 1 - n_{0y}, n_{0x})).
    \end{split}
\end{equation}
By dividing this number by the total number of pairs $\bpm n \\ 2 \epm$, we get the approximate upper bound for the TB case:
	\begin{equation}
	    \begin{split}\nonumber
	        	| \tau(\bx,\by) | & \lessapprox 2\max(\pi_{0y}, 1 - \pi_{0y}) \{1 - \max(\pi_{0y}, 1 - \pi_{0y}, \pi_{0x})\}.
	    \end{split}
	\end{equation}

\section{Proofs of the Theorems 3 and 4}\label{sec:proof}	


\textbf{\textit {Proof of Theorem~\ref{thm:BCbound}.}} From \citet{Alan1988},
\begin{equation}
    \begin{split}\label{eq:errbd-Alan}
        |F^{-1}(\tau, \Delta) -\widetilde F^{-1}(\tau, \Delta)| & \le \dfrac{d}{8}h^2 \sup_{x = \tau, \Delta} \left|\dfrac{\partial^2 F^{-1}(x)}{\partial x^2}\right|
    \end{split}
\end{equation}
holds with $d=2$, where $h$ is the maximal grid width. Lemmas~\ref{lem:bc-2nd-tau} and \ref{lem:bc-2nd-delta} give the upper bounds of second derivatives:
\begin{equation}
\begin{split}\nonumber
    \left|\frac{\partial^2 F^{-1}}{\partial \tau^2} \right| \leq |r|\frac{\pi^2}{2}(2M^2 + 1)\exp(M^2) \quad \text{and} \quad
     \left|\frac{\partial^2 F^{-1}}{\partial \Delta^2}\right| \leq |r|\left\{1 + \frac{\sqrt{\pi}}{\sqrt{2}}M\exp\left(\frac{M^2}{2}\right)\right\}.
\end{split}
\end{equation}
The upper bound of the second derivative with respect to $\tau$ is always larger than the one with respect to $\Delta$ for all non-negative $M$. Thus, plugging in the second derivative with respect to $\tau$ into \eqref{eq:errbd-Alan} and using the constant $\frac{d}{8}\frac{\pi^2}{2} \le 2$ concludes the proof of Theorem 3.

\hfill \qedsymbol

\noindent\textbf{\textit {Proof of Theorem~\ref{thm:TCbound}.}} Similarly to the proof of Theorem 3, assuming that $\Delta \le M$ for a positive $M$, we have the upper bounds of second derivatives with respect to $\tau$ and $\Delta$ from Lemma~\ref{lem:tc-2nd-tau} and \ref{lem:tc-2nd-delta}:
\begin{equation}
    \begin{split}\nonumber
        \left| \frac{\partial^2 F^{-1}(\tau, \Delta)}{\partial \tau^2} \right| & \le   \dfrac{16|r|}{\left\{\Phi(-\sqrt{2}M)\right\}^3},\\
        \left| \frac{\partial^2 F^{-1}}{\partial \Delta^2}  \right|
        & \le \dfrac{\sqrt{1-r^2}\left( 4 + 6 M \right) }{\Phi(-\sqrt{2}M)}  + \dfrac{5\sqrt{1-r^2} }{\left\{ \Phi(-\sqrt{2}M)\right\}^2}. \\
    \end{split}
\end{equation}
%
Since $M$ is a positive and finite number, $\Phi(-\sqrt{2}M) \in (0, 0.5)$. For example, if $M = 1.64$, $\Phi(-\sqrt{2}M) \approx 0.01019$. Between these two upper bounds of the second derivatives, which one is larger depends on $\Phi(-\sqrt{2}M)$ and the size of latent correlation $r$. Thus we obtain
\begin{equation}
    \begin{split}\nonumber
        \sup_{x = \tau, \Delta} \left|\dfrac{\partial^2 F^{-1}(x)}{\partial x^2}\right| & \le \dfrac{16}{\left\{\Phi(-\sqrt{2}M)\right\}^2} \max \left( \dfrac{|F^{-1}(\tau, \Delta)|}{\Phi(-\sqrt{2}M)},   \sqrt{1-\left\{F^{-1}(\tau, \Delta)\right\}^2} \right).
    \end{split}
\end{equation}


\hfill \qedsymbol

\section{Supporting lemmas}
\begin{lemma}\label{lem:bc-2nd-tau}
Let $F^{-1}(\tau, \Delta)$ be the inverse bridge function for the binary/continuous case. Assume $-M \le  \Delta \le  M$ for some positive value of $M$, then
$$
\left|\frac{\partial^2 F^{-1}}{\partial \tau^2} \right| \leq \frac{\pi^2}{2}|r|(2M^2+1)\exp(M^2).
$$
\end{lemma}

\noindent \textbf{\textit {Proof of Lemma~\ref{lem:bc-2nd-tau}.}}
The bridge function for binary/continuous case is
\begin{equation}\nonumber
\tau = F_{BC}(r, \Delta)= F(r, \Delta) = 4\Phi_2(\Delta, 0; r/\sqrt{2}), - 2 \Phi(\Delta)
\end{equation}
with its inverse $r = F^{-1}(\tau, \Delta)$. We first calculate the  partial derivatives of bridge function itself, and then use Lemmas~\ref{lem:inv_diff1} and \ref{lem:inv_diff2} to find the second partial derivatives of bridge inverse function.

Consider the first partial derivative of bridge function with respect to $r$. Using Lemma~S1 from \citet{mixedCCA}:
\begin{equation}
\begin{split}\label{eq:bc-1st_r}
    \frac{\partial F(r, \Delta)}{\partial r} &= 4\frac{\partial \Phi_2(\Delta, 0; r/\sqrt{2})}{\partial r} = 4 \phi_2(\Delta, 0; r/\sqrt{2})/\sqrt{2} = 2\sqrt{2}\phi_2(\Delta, 0; r/\sqrt{2}).
\end{split}
\end{equation}
 Using Lemma~\ref{lem:2dnormal_deriv} and the chain rule, the second partial derivative of bridge function with respect to $r$ is
\begin{equation}
\begin{split}\label{eq:bc-2nd_r}
\frac{\partial^2 F(r, \Delta)}{\partial r^2} &= 2\sqrt{2} \frac{\partial \phi_2(\Delta, 0; r/\sqrt{2})}{\partial (r/\sqrt{2})} \dfrac{1}{\sqrt{2}}  = 2  \phi_2(\Delta, 0; r/\sqrt{2}) \left[\frac{r/\sqrt{2}}{1 - r^2/2} - \frac{\Delta^2 r/\sqrt{2}}{(1-r^2/2)^2}\right]\\
&= \phi_2(\Delta, 0; r/\sqrt{2})\frac{\sqrt{2}r}{(1-r^2/2)^2}\left[1-r^2/2 - \Delta^2\right]\\
& =  \phi_2(-\Delta, 0; r/\sqrt{2})  \dfrac{2\sqrt{2}r}{(2-r^2)^2} (2-r^2-2\Delta^2).
\end{split}
\end{equation}
From Lemma~\ref{lem:inv_diff2}, we have
\begin{equation}\label{eq:bridgeinv_2nd_tau}
    \frac{\partial^2 F^{-1}(\tau, \Delta)}{\partial \tau^2} =- \frac{\partial^2 F(r, \Delta)}{\partial r^2} \left(\frac1{ \frac{\partial F(r, \Delta)}{\partial r}}\right)^3.
\end{equation}
Therefore, plugging \eqref{eq:bc-1st_r} and \eqref{eq:bc-2nd_r} into \eqref{eq:bridgeinv_2nd_tau} gives
\begin{equation}
\begin{split}\nonumber
    \frac{\partial^2 F^{-1}(\tau, \Delta)}{\partial \tau^2} 
    & = - \phi_2(-\Delta, 0; r/\sqrt{2})  \dfrac{2\sqrt{2}r}{(2-r^2)^2} (2-r^2-2\Delta^2)  \left\{\frac1{2\sqrt{2}\phi_2(\Delta, 0; r/\sqrt{2})}\right\}^3\\
    & = - \dfrac{r}{(2-r^2)^2} (2-r^2-2\Delta^2)  \left\{\frac1{2\sqrt{2}\phi_2(\Delta, 0; r/\sqrt{2})}\right\}^2.
\end{split}
\end{equation}
We further simplify using $\phi_2\left(\Delta, 0; r/\sqrt{2}\right) = \dfrac{1}{\pi\sqrt{2(2-r^2)}} \exp \left\{ -\dfrac{\Delta^2}{2-r^2}\right\}$.
\begin{equation}
\begin{split}\nonumber
    \frac{\partial^2 F^{-1}(\tau, \Delta)}{\partial \tau^2} &=
    - \dfrac{r}{(2-r^2)^2} (2-r^2-2\Delta^2)  \frac1{8} \pi^2 2(2-r^2) \exp \left\{\dfrac{\Delta^2}{2-r^2}\right\} \\
    & = - \dfrac{\pi^2}{4} \dfrac{r}{2-r^2} (2-r^2-2\Delta^2) \exp \left\{\dfrac{\Delta^2}{2-r^2}\right\}\\
    & = - \dfrac{\pi^2 r}{4} (1-\dfrac{2\Delta^2}{2-r^2}) \exp \left\{\dfrac{\Delta^2}{2-r^2}\right\}.
\end{split}
\end{equation}
Let $z = \Delta^2/(2-r^2)$. Since $2-r^2 \in [1,2]$, we have $\Delta^2/2 \leq z \leq \Delta^2$ regardless of the value of $r$. Then, the second derivative with respect to $\tau$ is
\begin{equation}\label{eq:bc-2nd-tau-intermsof-z}
\frac{\partial^2 F^{-1}}{\partial \tau^2} = \frac{\pi^2 r}2(2z-1)\exp(z).
\end{equation}
For $z\leq \frac{1}{2}$, we have the second derivative bounded by
\begin{equation}\label{eq:z_le0.5}
\left|\frac{\partial^2 F^{-1}}{\partial \tau^2} \right| \leq \frac{\pi^2  |r|}{2}.
\end{equation}
For $z > \frac{1}{2}$, \eqref{eq:bc-2nd-tau-intermsof-z} is strictly increasing in $z$. Thus, if $|\Delta|\leq M$, we have
\begin{equation}\label{eq:z_ge0.5}
\left|\frac{\partial^2 F^{-1}}{\partial \tau^2} \right| \leq \frac{\pi^2}{2}|r(2M^2-1)|\exp(M^2).
\end{equation}
To combine both cases \eqref{eq:z_le0.5} and \eqref{eq:z_ge0.5} for all $z$, we use $\max \left( 1, |2M^2-1|  \right) \le |2M^2 + 1|$. 
\hfill \qedsymbol



\begin{lemma}\label{lem:bc-2nd-delta}
Let $F^{-1}(\tau, \Delta)$ be the inverse bridge function for the binary/continuous case. Assume $-M \le  \Delta \le  M$ for some positive value of $M$, then
$$
\left|\frac{\partial^2 F^{-1}}{\partial \Delta^2} \right| \leq |r|\left(1 + \frac{\sqrt{\pi}}{\sqrt{2}}M\exp(M^2/2)\right).
$$
\end{lemma}

\noindent \textbf{\textit {Proof of Lemma~\ref{lem:bc-2nd-delta}.}} First, we calculate the first partial derivative of bridge function with respect to $\Delta$:
\begin{equation}
    \begin{split}\label{eq:bc-1st-delta}
        \frac{\partial F(r, \Delta)}{\partial \Delta} &= 4\int_{-\infty}^0\phi_2(\Delta, x_2; r/\sqrt{2})dx_2 - 2\phi(\Delta)\\
        & =  4\int_{-\infty}^0\phi(\Delta)\phi(x_2|x_1=\Delta, r/\sqrt{2})dx_2 - 2\phi(\Delta)\\
        &= 2\phi(\Delta)\left\{2\Phi(0|x_1=\Delta, r/\sqrt{2}) - 1\right\}, 
    \end{split}
\end{equation}
where $\phi(x_2|x_1=\Delta, r/\sqrt{2})$ denotes the conditional distribution of $X_2|X_1 = \Delta$ where the correlation between $X_1$ and $X_2$ is $r/\sqrt{2}$. For bivariate random variable $(X_1, X_2)$ with correlation $r/\sqrt{2}$, $\phi(x_2|x_1=\Delta, r/\sqrt{2})$ is the density of the normal distribution with mean $r\Delta/\sqrt{2}$ and variance $1-r^2/2$. Thus, $\Phi(0|x_1=\Delta, r/\sqrt{2})$ can be simplified as below where $z = \frac{-r\Delta/\sqrt{2}}{\sqrt{1-r^2/2}}$.
\begin{equation}
    \begin{split}\label{eq:bc-1st-delta-part}
    \Phi(0|x_1=\Delta, r/\sqrt{2}) &= \int_{-\infty}^{0} \dfrac{1}{\sqrt{2\pi \left(1-\frac{r^2}{2} \right)}}\exp\left\{ -\dfrac{1}{2}\left(\dfrac{x_2 - \frac{r\Delta}{\sqrt{2}}}{\sqrt{1-\frac{r^2}{2}}}\right)^2 \right\} dx_2 = \Phi\left(z\right).
    \end{split}
\end{equation}
Plugging \eqref{eq:bc-1st-delta-part} into \eqref{eq:bc-1st-delta} gives
\begin{equation}\label{eq:bc-1st-delta-withz}
    \frac{\partial F(r, \Delta)}{\partial \Delta} = 2\phi(\Delta)\left\{2\Phi(z) - 1\right\}.
\end{equation}
Based on the chain rule, the first partial derivative of inverse bridge function with respect to $\Delta$ is
\begin{equation}
    \begin{split}\label{eq:inverse-1st-delta}
        \frac{\partial F^{-1} (\tau, \Delta)}{\partial \Delta} = \frac{\partial F^{-1} (\tau, \Delta) }{\partial \tau}  \frac{\partial \tau }{\partial \Delta} = \frac1{\frac{\partial F (r, \Delta) }{\partial r}} \frac{\partial F (r, \Delta) }{\partial \Delta}.
    \end{split}
\end{equation}
Using \eqref{eq:bc-1st-delta-withz} and \eqref{eq:bc-1st_r}, we obtain
\begin{equation}
    \begin{split}\nonumber
    \frac{\partial F^{-1}}{\partial \Delta} =\frac{\partial F}{\partial \Delta}\Big/ \frac{\partial F}{\partial r} = \frac{2\phi(\Delta)\left\{2\Phi(z) - 1\right\}}{2\sqrt{2}\phi_2(\Delta, 0; r/\sqrt{2})}.
    \end{split}
\end{equation}
In addition, we replace $\phi$ and $\phi_2$ with the normal density function formula and rearrange further to simplify.
\begin{equation}
    \begin{split}\nonumber
    \frac{\partial F^{-1}}{\partial \Delta}
    & = \frac{\frac1{\sqrt{2\pi}} \exp\{-\Delta^2/2\} \left\{2\Phi(z) - 1\right\}}{\sqrt{2} \frac1{\pi \sqrt{2(2-r^2)}} \exp\{-\frac{\Delta^2}{2-r^2}\} }\\
    & = \dfrac{\pi \sqrt{2} \sqrt{2-r^2}}{2\sqrt{\pi}} \exp\left\{ -\frac{\Delta^2(2-r^2)  - 2\Delta^2}{2(2-r^2)}\right\} \left\{2\Phi(z) - 1\right\}\\
    & = \left(\sqrt{\pi} \sqrt{1-r^2/2}\right)\exp\left\{-\frac1{2} \left(\frac{r\Delta/\sqrt{2}}{\sqrt{1-r^2/2}} \right)^2\right\} \left\{2\Phi(z) - 1\right\}\\
    & = \frac{\sqrt{1-r^2/2} \left\{2\Phi(z) - 1 \right\}}{\sqrt{2}\phi(z)}.
    \end{split}
\end{equation}
Next, the second derivative  with respect to $\Delta$ is
\begin{equation}\nonumber
     \frac{\partial^2 F^{-1}}{\partial \Delta^2} = \frac{\sqrt{1-r^2/2}}{\sqrt{2}} \left\{ \dfrac{2 \frac{\partial\Phi(z)}{\partial \Delta}}{\phi(z)} - \dfrac{ (2\Phi(z) - 1)}{\phi(z)^2}\frac{\partial \phi(z)}{\partial \Delta} \right\}.
\end{equation}
Using the chain rules:
\begin{equation}\nonumber
     \frac{\partial \Phi(z)}{\partial \Delta} = \frac{\partial \Phi(z)}{\partial z} \frac{\partial z}{\partial \Delta} = \phi(z) \frac{\partial z}{\partial \Delta}
     \quad \text{ and } \quad 
     \frac{\partial \phi(z)}{\partial \Delta} = \frac{\partial \phi(z)}{\partial z} \frac{\partial z}{\partial \Delta} = -z\phi(z) \frac{\partial z}{\partial \Delta}, 
\end{equation}
we get
\begin{equation}
    \begin{split}\nonumber
     \frac{\partial^2 F^{-1}}{\partial \Delta^2}
     &= \frac{\sqrt{1-r^2/2}}{\sqrt{2}} \left\{ \dfrac{2 \phi(z) \frac{\partial z}{\partial \Delta}}{\phi(z)} +  \dfrac{ (2\Phi(z) - 1)}{\phi(z)^2} z\phi(z) \frac{\partial z}{\partial \Delta} \right\} \\
     &= \frac{\sqrt{1-r^2/2}}{\sqrt{2}} \left\{ 2\frac{\partial z}{\partial \Delta} +  \dfrac{ z(2\Phi(z) - 1)}{\phi(z)}\frac{\partial z}{\partial \Delta} \right\}.
\end{split}
\end{equation}
Plugging $\dfrac{\partial z}{ \partial \Delta} = -\dfrac{r/\sqrt{2}}{\sqrt{1-r^2/2}}$ into the previous equation gives
\begin{equation}
\begin{split}\nonumber
     \frac{\partial^2 F^{-1}}{\partial \Delta^2}&=\frac{\sqrt{1-r^2/2}}{\sqrt{2}}2 \frac{-r/\sqrt{2}}{\sqrt{1-r^2/2}} +  \frac{\sqrt{1-r^2/2}}{\sqrt{2}}\frac{ z\frac{-r/\sqrt{2}}{\sqrt{1-r^2/2}}(2\Phi(z) - 1)}{\phi(z)}\\
     & = -r +\frac{-rz(2\Phi(z) - 1)}{2\phi(z)} = -r \left\{1 + \frac{z(2\Phi(z)-1)}{2\phi(z)}\right\}.    
\end{split}
\end{equation}
Using $| 2\Phi(z)-1 | \le 1$ and $1/\phi(z) = \sqrt{2\pi} \exp(z^2/2)$, we find that the second derivative with respect to $\Delta$ is bounded by
\begin{equation}\nonumber
 \left|\frac{\partial^2 F^{-1}}{\partial \Delta^2}\right| \leq |r|\left(1 + \frac{\sqrt{\pi}}{\sqrt{2}}|z|\exp(z^2/2)\right).
\end{equation}
Recall that $z = \frac{-r\Delta/\sqrt{2}}{\sqrt{1-r^2/2}}$. $|z| \le |\Delta|$ follows from $-M \le  \Delta \le  M$ for some positive constant $M$ and $-1 \le r \le 1$. This completes the proof.

\hfill \qedsymbol

\begin{lemma}\label{lem:tc-2nd-tau}
Let $F^{-1}(\tau, \Delta)$ be the inverse bridge function for the truncated/continuous case. Assume $\Delta \le  M$ for some positive value of $M$, then
$$
\left|\frac{\partial^2 F^{-1}}{\partial \tau^2} \right| \le \dfrac{16|r|}{\left\{\Phi(-\sqrt{2}M)\right\}^3}.
$$
\end{lemma}

\noindent \textbf{\textit {Proof of Lemma~\ref{lem:tc-2nd-tau}.}}
The bridge function for truncated/continuous case is
\begin{equation}\nonumber
\tau = F_{TC}(r, \Delta)= F(r, \Delta) = -2 \Phi_2 (-\Delta,0; 1/\sqrt{2} ) +4\Phi_3 \left(-\Delta,0,0; \bSigma_3(r)\right)
\end{equation}
where 
$\bSigma_3(r) = \bpm
        1 & 1/\sqrt{2} & r/\sqrt{2}\\
        1/\sqrt{2} & 1 & r\\
        r/\sqrt{2} & r & 1
    \epm$
and we let inverse be $r = F^{-1}(\tau, \Delta)$. Similarly to the proof of Lemma~\ref{lem:bc-2nd-tau}, we first calculate the first partial derivative of bridge function with respect to $r$.
\begin{equation}
\begin{split}\label{eq:tc_first_r}
    \frac{\partial F(r, \Delta)}{\partial r} & = 4  \dfrac{\partial}{\partial r}\Phi_3 \left(-\Delta,0,0; \bSigma_3(r)\right)\\
    & = 4 \int_{-\infty}^{0}\phi_{3} (-\Delta, x_2, 0; \bSigma_3(r)) dx_2 \dfrac{1}{\sqrt{2}} + 4 \int_{-\infty}^{-\Delta}\phi_{3} (x_1, 0, 0; \bSigma_3(r)) dx_1.
\end{split}
\end{equation}
For $X_1, X_2, X_3 \sim N \left( \left( 0 ~ 0 ~ 0 \right)^\top, \bSigma_3(r)\right)$, the conditional distribution of $X_2|X_1= -\Delta, X_3 = 0$ is  normal distribution with mean $\dfrac{-\sqrt{2}(1-r^2)\Delta}{2-r^2}$ and the variance $\dfrac{1-r^2}{2-r^2}$. Therefore, the first integral term of \eqref{eq:tc_first_r} can be simplified as
\begin{equation}
\begin{split}\label{eq:tc_first_r_part1}
\int_{-\infty}^{0}\phi_{3} (-\Delta, x_2, 0; \bSigma_3(r)) dx_2 & = \int_{-\infty}^{0} \phi_{2} (-\Delta, 0; r/\sqrt{2}) \phi \left(x_2| x_1 = -\Delta, x_3 = 0; \dfrac{1-r^2}{2-r^2}\right) dx_2\\
& = \phi_{2} (-\Delta, 0; r/\sqrt{2}) \Phi\left(\Delta \sqrt{\dfrac{2(1-r^2)}{2-r^2}}\right).
\end{split}
\end{equation}
Similarly, using the conditional distribution $X_1|X_2 = 0, X_3 = 0 \sim N\left(0, 1/2\right)$, the second term of \eqref{eq:tc_first_r} can be re-written as
\begin{equation}
\begin{split}\label{eq:tc_first_r_part2}
\int_{-\infty}^{-\Delta}\phi_{3} (x_1, 0, 0; \bSigma_3(r)) dx_1 & = \int_{-\infty}^{-\Delta} \phi_{2} (0, 0; r) \phi (x_1| x_2 = 0, x_3 = 0; 1/2) dx_1\\
& = \phi_{2} (0, 0; r)\Phi(-\sqrt{2}\Delta).
\end{split}
\end{equation}
Plugging \eqref{eq:tc_first_r_part1} and \eqref{eq:tc_first_r_part2} into \eqref{eq:tc_first_r} results in
\begin{equation}\nonumber
\frac{\partial F(r, \Delta)}{\partial r} = 2\sqrt{2}  \phi_{2} \left(-\Delta, 0; r/\sqrt{2}\right) \Phi\left(\Delta \sqrt{\frac{2(1-r^2)}{2-r^2}}\right) + 4 \phi_{2} (0, 0; r) \Phi(-\sqrt{2}\Delta).
\end{equation}
Using the fact that $\phi_2(0,0;r) = \frac1{2\pi\sqrt{1-r^2}}$ further gives
\begin{equation}\label{eq:tc_first_r_final}
    \frac{\partial F(r, \Delta)}{\partial r} = 2\sqrt{2} \phi_{2} \left(-\Delta, 0; r/\sqrt{2}\right) \Phi\left(\Delta \sqrt{\frac{2(1-r^2)}{2-r^2}}\right) + \frac2{\pi\sqrt{1-r^2}}\Phi(-\sqrt{2}\Delta).
\end{equation}
Then, the second partial derivative with respect to $r$ is
\begin{equation}\label{eq:tc_second_r}
    \begin{split}
\dfrac{\partial^2 F(r, \Delta)}{\partial r^2}  & = \dfrac{\partial}{\partial r} \left\{ 2\sqrt{2}  \phi_{2} \left(-\Delta, 0; r/\sqrt{2}\right) \Phi\left(\Delta \sqrt{\dfrac{2(1-r^2)}{2-r^2}}\right) +  \dfrac{2}{\pi\sqrt{1-r^2}} \Phi(-\sqrt{2}\Delta) \right\}\\
    & = 2\sqrt{2}  \dfrac{\partial}{\partial r} \phi_{2} \left(-\Delta, 0; r/\sqrt{2}\right) \Phi\left(\Delta \sqrt{\dfrac{2(1-r^2)}{2-r^2}}\right) \\
    & \quad + 2\sqrt{2}  \phi_{2} \left(-\Delta, 0; r/\sqrt{2}\right) \dfrac{\partial}{\partial r} \Phi\left(\Delta \sqrt{\dfrac{2(1-r^2)}{2-r^2}}\right)\\
        & \quad +  \dfrac{2}{\pi}\Phi(-\sqrt{2}\Delta) \dfrac{\partial}{\partial r} \left(\frac1{\sqrt{1-r^2}}\right).
    \end{split}
\end{equation}
Next we find three partial derivatives with respect to $r$ in the previous display separately.
\begin{equation}
    \begin{split}\label{eq:tc_second_r_part1}
        \dfrac{\partial}{\partial r} \phi_{2} \left(-\Delta, 0; r/\sqrt{2}\right) & = \dfrac{1}{\sqrt{2}} \phi_2(-\Delta, 0; r/\sqrt{2}) \left[\frac{r/\sqrt{2}}{1 - r^2/2} - \frac{\Delta^2 r/\sqrt{2}}{(1-r^2/2)^2}\right]\\
        & = \phi_2(-\Delta, 0; r/\sqrt{2}) \left[\frac{r}{2 - r^2} - \frac{\Delta^2 r}{(2-r^2)^2/2}\right]\\
        & = \phi_2(-\Delta, 0; r/\sqrt{2}) \left[\frac{r (2 - r^2) - 2\Delta^2 r}{(2-r^2)^2}\right]\\
        & = \phi_2(-\Delta, 0; r/\sqrt{2})  \dfrac{r(2-r^2-2\Delta^2)}{(2-r^2)^2}.
    \end{split}
\end{equation}
\begin{equation}
    \begin{split}\label{eq:tc_second_r_part2}
        \dfrac{\partial}{\partial r} \Phi\left(\Delta \sqrt{\dfrac{2(1-r^2)}{2-r^2}}\right) & = \phi\left(\Delta \sqrt{\dfrac{2(1-r^2)}{2-r^2}}\right) \dfrac{\partial}{\partial r} \left\{\Delta \sqrt{\dfrac{2(1-r^2)}{2-r^2}} \right\}\\
        & = \phi\left(\Delta \sqrt{\dfrac{2(1-r^2)}{2-r^2}}\right) \dfrac{\sqrt{2}\Delta}{2} \left(\frac{1-r^2}{2-r^2}\right)^{-1/2} \dfrac{-2r}{(2-r^2)^2}\\
        & = \phi\left(\Delta \sqrt{\dfrac{2(1-r^2)}{2-r^2}}\right) 
        \dfrac{-\sqrt{2}r\Delta}{(1-r^2)^{1/2} (2-r^2)^{3/2} }.
    \end{split}
\end{equation}
\begin{equation}
    \begin{split}\label{eq:tc_second_r_part3}
        \dfrac{\partial}{\partial r} \left(\frac1{\sqrt{1-r^2}}\right) & = \left(-\dfrac{1}{2}\right) (1-r^2)^{-3/2} (-2r) = \dfrac{r}{ (1-r^2)^{3/2}}.
    \end{split}
\end{equation}
Plugging \eqref{eq:tc_second_r_part1}, \eqref{eq:tc_second_r_part2} and \eqref{eq:tc_second_r_part3} into \eqref{eq:tc_second_r} and rearranging yields
\begin{equation}
\begin{split}\label{eq:tc_second_r_final}
    \dfrac{\partial^2 F(r, \Delta)}{\partial r^2}  & = 2\sqrt{2}  \phi_2(-\Delta, 0; r/\sqrt{2})  \dfrac{r(2-r^2-2\Delta^2)}{(2-r^2)^2} \Phi\left(\Delta \sqrt{\dfrac{2(1-r^2)}{2-r^2}}\right) \\
    & \quad + 2\sqrt{2}  \phi_{2} \left(-\Delta, 0; r/\sqrt{2}\right) \phi\left(\Delta \sqrt{\dfrac{2(1-r^2)}{2-r^2}}\right) 
        \dfrac{-\sqrt{2}r\Delta}{(1-r^2)^{1/2} (2-r^2)^{3/2}}\\
        & \quad +  \dfrac{2r}{ \pi (1-r^2)^{3/2}} \Phi(-\sqrt{2}\Delta)\\
    & = - \frac{2\sqrt{2}r(2-r^2-2\Delta^2) }{(2-r^2)^2} \phi_2(-\Delta, 0; r/\sqrt{2}) \Phi\left(\Delta \sqrt{\frac{2(1-r^2)}{2-r^2}}\right) \\
    & \quad - \frac{4r\Delta }{(2-r^2)^{3/2} (1-r^2)^{1/2}} \phi_2(-\Delta, 0; r/\sqrt{2}) \phi\left(\Delta \sqrt{\frac{2(1-r^2)}{2-r^2}}\right) \\
    & \quad + \dfrac{2r}{\pi \sqrt{(1-r^2)^{3}}} \Phi(-\sqrt{2}\Delta).
\end{split}
\end{equation}
From \eqref{eq:bridgeinv_2nd_tau}, we have
\begin{equation}\nonumber
    \frac{\partial^2 F^{-1}(\tau, \Delta)}{\partial \tau^2} = - \dfrac{ \frac{\partial^2 F(r, \Delta)}{\partial r^2} }{\left\{ 2\sqrt{2} \phi_{2} \left(-\Delta, 0; r/\sqrt{2}\right) \Phi\left(\Delta \sqrt{\frac{2(1-r^2)}{2-r^2}}\right) + \frac2{\pi\sqrt{1-r^2}}\Phi(-\sqrt{2}\Delta) \right\}^3}.
\end{equation}
Here, $\phi_{2} \left(-\Delta, 0; r/\sqrt{2}\right) \Phi\left(\Delta \sqrt{\frac{2(1-r^2)}{2-r^2}}\right)$ in the denominator is always non-negative. We find the upper bound of this second derivative of $F^{-1}$ with respect to $\tau$ by eliminating the non-negative term in the denominator.
\begin{equation}
    \begin{split}\nonumber
        \left| \frac{\partial^2 F^{-1}(\tau, \Delta)}{\partial \tau^2} \right|
        \le \left|  \dfrac{ \frac{\partial^2 F(r, \Delta)}{\partial r^2} }{ \left\{ \frac2{\pi\sqrt{1-r^2}}\Phi(-\sqrt{2}\Delta) \right\}^3 } \right|.
    \end{split}
\end{equation}
Plugging \eqref{eq:tc_second_r_final} into the numerator of the previous display and rearranging results in
\begin{equation}
    \begin{split}\nonumber
        & \left| \frac{\partial^2 F^{-1}(\tau, \Delta)}{\partial \tau^2} \right| \\
        & \le \left|  - \frac{2\sqrt{2}r(2-r^2-2\Delta^2) }{(2-r^2)^2} \phi_2(-\Delta, 0; r/\sqrt{2}) \Phi\left(\Delta \sqrt{\frac{2(1-r^2)}{2-r^2}}\right) \dfrac{\pi^3 (1-r^2)^{3/2}}{2^3 \left\{\Phi(-\sqrt{2}\Delta)\right\}^3} \right.\\
        & \quad \quad + \frac{4r\Delta }{(2-r^2)^{3/2} (1-r^2)^{1/2}} \phi_2(-\Delta, 0; r/\sqrt{2}) \phi\left(\Delta \sqrt{\frac{2(1-r^2)}{2-r^2}}\right) \dfrac{\pi^3 (1-r^2)^{3/2}}{2^3 \left\{\Phi(-\sqrt{2}\Delta)\right\}^3}\\
        & \quad \quad - \left. \frac{r\pi^2 }{\left(\Phi(-\sqrt{2}\Delta)\right)^2}  \right|\\
        & \le |r| \left|\frac{2\sqrt{2}(2-r^2-2\Delta^2) }{(2-r^2)^2} \phi_2(-\Delta, 0; r/\sqrt{2}) \Phi\left(\Delta \sqrt{\frac{2(1-r^2)}{2-r^2}}\right) \dfrac{\pi^3 (1-r^2)^{3/2}}{2^3 \left\{\Phi(-\sqrt{2}\Delta)\right\}^3} \right|\\
        & \quad \quad + |r| \left|\frac{4\Delta (1-r^2) }{(2-r^2)^{3/2} } \phi_2(-\Delta, 0; r/\sqrt{2}) \phi\left(\Delta \sqrt{\frac{2(1-r^2)}{2-r^2}}\right) \dfrac{\pi^3}{2^3 \left\{\Phi(-\sqrt{2}\Delta)\right\}^3} \right|\\
        & \quad \quad + |r| \left| \frac{\pi^2 }{\left(\Phi(-\sqrt{2}\Delta)\right)^2}  \right|.
    \end{split}
\end{equation}
To further simplify, we use $2-r^2 \in [1,2]$, $1-r^2 \in [0,1]$, $\Phi(x) \le 1$ for all $x\in \mathbb{R}$ and $\frac{1-r^2}{(2-r^2)^{3/2}} \le \frac{1}{2\sqrt{2}}$ for all $r \in [-1, 1]$. 
In addition, we have $$\phi\left(\Delta \sqrt{\frac{2(1-r^2)}{2-r^2}}\right) \le \phi\left(|\Delta|\right) = \dfrac{1}{\sqrt{2\pi}}\exp \left\{ -\dfrac{\Delta^2}{2} \right\} $$ from $0 \le \frac{2(1-r^2)}{2-r^2} \le 1$ and
\begin{equation}\nonumber
        \phi_2\left(-\Delta, 0; r/\sqrt{2}\right) = \dfrac{1}{\pi\sqrt{2(2-r^2)}} \exp \left\{ -\dfrac{\Delta^2}{2-r^2}\right\} \le \dfrac{1}{\pi\sqrt{2}} \exp \left\{ -\dfrac{\Delta^2}{2}\right\}.
\end{equation}
Therefore,
\begin{equation}
    \begin{split}\nonumber
        & \left| \frac{\partial^2 F^{-1}(\tau, \Delta)}{\partial \tau^2} \right| \\
        & \le |r| \left| \dfrac{2\sqrt{2}}{\pi\sqrt{2}} (2-r^2 - 2\Delta^2) \exp \left\{ -\dfrac{\Delta^2}{2}\right\}  \dfrac{\pi^3}{2^3 \left\{\Phi(-\sqrt{2}\Delta)\right\}^3} \right|\\
        & \quad + |r| \left| 4 \dfrac{|\Delta|}{\pi\sqrt{2}} \exp \left\{ -\dfrac{\Delta^2}{2}\right\} \dfrac{1}{\sqrt{2\pi}}\exp \left\{ -\dfrac{\Delta^2}{2} \right\} \dfrac{\pi^3}{2^3 \left\{\Phi(-\sqrt{2}\Delta)\right\}^3} \right|
        + |r| \left| \frac{\pi^2 }{\left(\Phi(-\sqrt{2}\Delta)\right)^2}  \right|,
    \end{split}
\end{equation}
and further cancelling and rearranging gives
\begin{equation}
    \begin{split}\nonumber
        & \left| \frac{\partial^2 F^{-1}(\tau, \Delta)}{\partial \tau^2} \right| \\
        & \le |r| \dfrac{\pi^2}{2^2} \dfrac{ | 2-r^2 - 2\Delta^2 |\exp \left\{ -\dfrac{\Delta^2}{2}\right\}}{\left\{\Phi(-\sqrt{2}\Delta)\right\}^3}
        + |r| \dfrac{\pi^{3/2}}{4} \dfrac{ |\Delta|\exp \left\{ -\Delta^2\right\} }{\left\{\Phi(-\sqrt{2}\Delta)\right\}^3}
        + |r| \frac{\pi^2 }{\left\{\Phi(-\sqrt{2}\Delta)\right\}^2}.
    \end{split}
\end{equation}
Assume that $\Delta \le M$ for some positive constant $M$. Otherwise, $\Phi(-\sqrt{2}\Delta) \to 0$ which leads the upper bound of second derivative of $F^{-1}$ with respect to $\tau$ to infinity. Using the fact that $\left| \Delta\exp\left\{-\Delta^2 \right\} \right| \le \exp\{-1/2\}/\sqrt{2}$ for all $\Delta \in \mathbb{R}$, we obtain
\begin{equation}
    \begin{split}\label{eq:tc-2nd-tau-interim}
        & \left| \frac{\partial^2 F^{-1}(\tau, \Delta)}{\partial \tau^2} \right| \\
        & \le |r| \dfrac{\pi^2}{2^2} \dfrac{ | 2-r^2 - 2\Delta^2 |\exp \left\{ -\dfrac{\Delta^2}{2}\right\}}{\left\{\Phi(-\sqrt{2}M)\right\}^3}
        + |r| \dfrac{\pi^{3/2}}{4\sqrt{2}} \dfrac{ \exp\{-1/2\} }{\left\{\Phi(-\sqrt{2}M)\right\}^3}
        + |r| \frac{\pi^2 }{\left\{\Phi(-\sqrt{2}M)\right\}^2}.
    \end{split}
\end{equation}
From $2-r^2 \in [1,2]$, we have $1-2\Delta^2 <2-r^2-2\Delta^2 < 2-2\Delta^2$. If $\Delta^2 = 3/4$, $ | 2-2\Delta^2 | = | 1 - 2\Delta^2| = 1/2$. For $\Delta^2 < 3/4$, $\left| 2-r^2-2\Delta^2 \right| \le 2-2\Delta^2$ which leads to  
\begin{equation}\label{eq:Dsq_le0.75}
    | 2-r^2 - 2\Delta^2| \exp \left\{ -\dfrac{\Delta^2}{2}\right\}  \le (2 - 2\Delta^2) \exp \left\{ -\dfrac{\Delta^2}{2}\right\} \le 2.
\end{equation}
On the other hand, if $\Delta^2 > 3/4$, $\left| 2-r^2-2\Delta^2 \right| \le 2\Delta^2 - 1$. Thus, 
\begin{equation}\nonumber
|2-r^2 - 2\Delta^2| \exp \left\{ -\dfrac{\Delta^2}{2}\right\}  \le  (2\Delta^2 - 1) \exp \left\{ -\dfrac{\Delta^2}{2}\right\}  \le  4 \exp \left\{ -1/2\right\}.
\end{equation}
Using $ 4 \exp \left\{ -1/2\right\} < 2$ and \eqref{eq:Dsq_le0.75}, we obtain $|2-r^2 - 2\Delta^2| \exp \left\{ -\Delta^2/2\right\}  \le  2$ for all $\Delta \in \mathbb{R}$ and \eqref{eq:tc-2nd-tau-interim} can be simplified as
\begin{equation}
    \begin{split}\nonumber
        \left| \frac{\partial^2 F^{-1}(\tau, \Delta)}{\partial \tau^2} \right| 
        \le  \dfrac{|r|\pi^2}{\left\{\Phi(-\sqrt{2}M)\right\}^3}  
        \left[ \dfrac{1}{2} 
        +  \dfrac{\exp\{-1/2\} }{4\sqrt{2\pi}} 
        + \left\{\Phi(-\sqrt{2}M)\right\}
        \right].
    \end{split}
\end{equation}
Finally, using $\pi^2\left\{ \dfrac{1}{2} + \dfrac{\exp(-1/2)}{ 4\sqrt{2\pi}} + 1 \right\} \le 16$ completes the proof. \hfill \qedsymbol

\begin{lemma}\label{lem:tc-2nd-delta}
Let $F^{-1}(\tau, \Delta)$ be the inverse bridge function for the truncated/continuous case. Assume $\Delta \le  M$ for some positive value of $M$, then
$$
\left|\frac{\partial^2 F^{-1}}{\partial \Delta^2} \right| \le \dfrac{\sqrt{1-r^2}\left( 4+6 M \right) }{\Phi(-\sqrt{2}M)}  + \dfrac{5\sqrt{1-r^2} }{\left\{ \Phi(-\sqrt{2}M)\right\}^2}.
$$
\end{lemma}

\noindent \textbf{\textit {Proof of Lemma~\ref{lem:tc-2nd-delta}.}}
Consider the first partial derivative of bridge function $F$ with respect to $\Delta$:
\begin{equation}
\begin{split}\label{eq:tc_first_delta}
    \dfrac{\partial F(r, \Delta)}{\partial \Delta} & = \dfrac{\partial}{\partial \Delta } \left[ -2 \Phi_2 (-\Delta,0; 1/\sqrt{2} ) +4\Phi_3 \left(-\Delta,0,0; \bSigma_3(r)\right) \right] \\
    & = -2 \dfrac{\partial}{\partial \Delta } \int_{-\infty}^{-\Delta} \int_{-\infty}^{0} \phi_2(x_1, x_2 ; 1/\sqrt{2} ) dx_2 dx_1\\
    & \quad + 4 \dfrac{\partial}{\partial \Delta }   \int_{-\infty}^{-\Delta} \int_{-\infty}^{0} \int_{-\infty}^{0} \phi_3 \left(x_1, x_2, x_3; \bSigma_3(r)\right) dx_3 dx_2 dx_1 \\
    & = 2 \int_{-\infty}^{0} \phi_2 (-\Delta, x_2; 1/\sqrt{2} ) dx_2 - 4 \int_{-\infty}^{0} \int_{-\infty}^{0} \phi_3 \left(-\Delta, x_2, x_3; \bSigma_3(r)\right) dx_2 dx_3
    \end{split}
\end{equation}
For a bivariate random variable $(X_1, X_2)$ with mean $\left( 0 ~ 0 \right)^\top$ and correlation $1/\sqrt{2}$, the conditional distribution $X_2 |X_1 = -\Delta$ satisfies $N(-\Delta/2, 1/2)$. Then, the first integral term in the previous display can be simplified as
\begin{equation}
    \begin{split}\label{eq:tc_first_delta_part1}
        \int_{-\infty}^{0} \phi_2 (-\Delta, x_2; 1/\sqrt{2} ) dx_2 & = 
        \int_{-\infty}^{0} \phi(-\Delta) \phi(x_2 | x_1 = -\Delta ; 1/\sqrt{2} ) dx_2\\
        & = \phi(\Delta) \int_{-\infty}^{0} \dfrac{1}{\sqrt{2\pi}/\sqrt{2}} \exp\left\{ -\dfrac{1}{2} \left(\dfrac{x_2 + \Delta/2}{1/\sqrt{2}}\right)^2 \right\} dx_2\\
        & = \phi(\Delta) \Phi(\Delta/\sqrt{2}).
    \end{split}
\end{equation}
Note that $\phi(\Delta) = \phi(-\Delta)$. For $X_1, X_2, X_3 \sim N \left( \left( 0~0~0\right)^\top, \bSigma_3(r)\right)$, since the conditional distribution is
\begin{equation}\label{eq:cond-23given1}
    X_2, X_3 |X_1 = -\Delta \sim N\left( \bpm -\frac{\Delta}{\sqrt{2}}\\ -\frac{r\Delta}{\sqrt{2}} \epm, \bSigma_2 (r) = \bpm 1/2 & r/2\\ r/2 & (2-r^2)/2 \epm\right),
\end{equation}
we have
\begin{equation}
\begin{split}\nonumber
    & \int_{-\infty}^{0} \int_{-\infty}^{0} \phi_3 \left(-\Delta, x_2, x_3; \bSigma_3(r)\right) dx_2 dx_3\\
    & \quad =\int_{-\infty}^{0} \int_{-\infty}^{0}\phi(-\Delta)  \phi_2 \left(x_2, x_3 | x_1 = -\Delta; \bSigma_2 (r) \right) dx_2 dx_3.
\end{split}
\end{equation}
If we let $g(r, \Delta)= \int_{-\infty}^{0} \int_{-\infty}^{0} \phi_2 \left(x_2, x_3 | x_1 = -\Delta; \bSigma_2 (r) \right) dx_2 dx_3$, then the second term in \eqref{eq:tc_first_delta} is
\begin{equation}\label{eq:tc_first_delta_part2}
    \int_{-\infty}^{0} \int_{-\infty}^{0} \phi_3 \left(-\Delta, x_2, x_3; \bSigma_3(r)\right) dx_2 dx_3 = \phi(\Delta)   g(r, \Delta).
\end{equation}
Using \eqref{eq:tc_first_delta_part1} and \eqref{eq:tc_first_delta_part2}, we obtain
\begin{equation}
\begin{split}\nonumber
    \dfrac{\partial F(r, \Delta)}{\partial \Delta}
    & = 2 \phi(\Delta) \Phi(\Delta/\sqrt{2}) - 4  \phi(\Delta)  g(r, \Delta).
\end{split}
\end{equation}
Using \eqref{eq:inverse-1st-delta}, we obtain the first partial derivative of $F^{-1}$ with respect to $\Delta$ 
\begin{equation}
\begin{split}\nonumber
        \frac{\partial F^{-1}}{\partial \Delta} &=\frac{\partial F}{\partial \Delta}\Big/ \frac{\partial F}{\partial r} = \frac{2 \phi(\Delta) \Phi(\Delta/\sqrt{2}) - 4  \phi(\Delta)  g(r, \Delta)}{2\sqrt{2} \phi_{2} \left(-\Delta, 0; r/\sqrt{2}\right) \Phi\left(\Delta \sqrt{\frac{2(1-r^2)}{2-r^2}}\right) + \frac2{\pi\sqrt{1-r^2}}\Phi(-\sqrt{2}\Delta)}.
\end{split}
\end{equation}
Let the whole term in the denominator of $\frac{\partial F^{-1}}{\partial \Delta}$ as $H(r, \Delta)$, then the second derivative  of $F^{-1}$ with respect to $\Delta$ is
\begin{equation}
\begin{split}\label{eq:tc_2nd_delta_step1}
        \frac{\partial^2 F^{-1}}{\partial \Delta^2} & = \frac{\partial }{\partial \Delta}\frac{2 \phi(\Delta) \Phi(\Delta/\sqrt{2}) - 4  \phi(\Delta)  g(r, \Delta)}{H(r, \Delta)}\\
        & = \frac{-2\Delta \phi(\Delta) \Phi(\Delta/\sqrt{2}) + \sqrt{2}\phi(\Delta)\phi(\Delta/\sqrt{2}) +  4\Delta \phi(\Delta) g(r, \Delta) -4 \phi(\Delta) \frac{\partial g(r, \Delta)}{\partial \Delta} }{H(r, \Delta)}\\
        & \quad - \frac{2 \phi(\Delta) \Phi(\Delta/\sqrt{2}) - 4  \phi(\Delta)  g(r, \Delta)}{H(r, \Delta)^2} \frac{\partial H(r, \Delta)}{\partial \Delta}.
\end{split}
\end{equation}
Here, $H(r, \Delta)$ can be rewritten as below using normal probability density function.
\begin{equation}\nonumber
        H(r, \Delta)  = \frac{2}{\pi \sqrt{2-r^2}} \exp\left\{-\frac{\Delta^2}{2-r^2}\right\} \Phi\left(\Delta \sqrt{\frac{2(1-r^2)}{2-r^2}}\right) + \frac2{\pi\sqrt{1-r^2}}\Phi(-\sqrt{2}\Delta)
\end{equation}
Since $\frac{2}{\pi \sqrt{2-r^2}} \exp\left\{-\frac{\Delta^2}{2-r^2}\right\} \Phi\left(\Delta \sqrt{\frac{2(1-r^2)}{2-r^2}}\right) \ge 0$, we have
\begin{equation}\nonumber
        H(r, \Delta) \ge \frac2{\pi\sqrt{1-r^2}}\Phi(-\sqrt{2}\Delta)
\end{equation}
and we find the upper bound for $H(r, \Delta)^{-1}$ and $H(r, \Delta)^{-2}$ from this lower bound.
\begin{equation}
    \begin{split}\label{eq:H_inv_ub}
        \frac{ 1 }{H(r, \Delta)}  & \le  \frac{\pi\sqrt{1-r^2}}{2\Phi(-\sqrt{2}\Delta)}\\
        \frac{ 1 }{H(r, \Delta)^2}  & \le  \frac{\pi^2(1-r^2)}{2^2\left\{\Phi(-\sqrt{2}\Delta)\right\}^2}.
    \end{split}
\end{equation}
The first derivative of $H(r, \Delta)$ is
\begin{equation}
    \begin{split}\nonumber
        \frac{\partial H(r, \Delta)}{\partial \Delta} & = \frac{2}{\pi \sqrt{2-r^2}} \exp\left\{-\frac{\Delta^2}{2-r^2}\right\} \left(-\frac{2\Delta}{2-r^2}\right) \Phi\left(\Delta \sqrt{\frac{2(1-r^2)}{2-r^2}}\right)\\
        & \quad + \frac{2}{\pi \sqrt{2-r^2}} \exp\left\{-\frac{\Delta^2}{2-r^2}\right\} \phi\left(\Delta \sqrt{\frac{2(1-r^2)}{2-r^2}}\right)\sqrt{\frac{2(1-r^2)}{2-r^2}} \\
        & \quad + \frac2{\pi\sqrt{1-r^2}} \phi(-\sqrt{2}\Delta) (-\sqrt{2}).
    \end{split}
\end{equation}
Using the facts that $| \Delta \exp \{ -\Delta^2/2\} | \le \exp \{ -1/2\}$, $\exp \{ -\Delta^2/2\} \le 1$ for all $\Delta$ and $\phi(x) \le \frac1{\sqrt{2\pi}}$ for all $x\in \mathbb{R}$, we obtain
\begin{equation}
    \begin{split}\label{eq:H_1stder_ub}
        \left| \frac{\partial H(r, \Delta)}{\partial \Delta}  \right| 
        & \le \frac{4}{\pi} \exp\left\{-\frac{1}{2}\right\}  + \frac{2}{\pi} \frac{1}{\sqrt{2\pi}} + \frac{2\sqrt{2}}{\pi\sqrt{1-r^2}} \frac{1}{\sqrt{2\pi}}\\
        & \le c_1 + c_2\frac1{\sqrt{1-r^2}}
    \end{split}
\end{equation}
where $c_1 = \frac{4}{\pi}\exp\left\{-\frac{1}{2}\right\} + \frac{\sqrt{2}}{\pi\sqrt{\pi}}$ and $c_2 = \frac{2}{\pi\sqrt{\pi}}$ are constants that do not depend on $r$ and $\Delta$. 
Plugging \eqref{eq:H_inv_ub} and \eqref{eq:H_1stder_ub} into \eqref{eq:tc_2nd_delta_step1} gives us
\begin{equation}
\begin{split}\nonumber
        & \left| \frac{\partial^2 F^{-1}}{\partial \Delta^2}  \right|\\
        & \quad \le  \frac{\pi\sqrt{1-r^2}}{2\Phi(-\sqrt{2}\Delta)} \left\{ 2|\Delta| \phi(\Delta) \Phi(\Delta/\sqrt{2}) + \dfrac{\sqrt{2}}{2\pi} +  4 |\Delta| \phi(\Delta) g(r, \Delta) + 4 \phi(\Delta) \left| \frac{\partial g(r, \Delta)}{\partial \Delta} \right| \right\} \\
        & \quad \quad - \left(c_1+ c_2\frac1{\sqrt{1-r^2}}\right) \frac{\pi^2(1-r^2)}{2^2\left\{\Phi(-\sqrt{2}\Delta)\right\}^2} \left| 2 \phi(\Delta) \Phi(\Delta/\sqrt{2}) - 4  \phi(\Delta)  g(r, \Delta) \right|.
\end{split}
\end{equation}
We use $|\Delta| \phi(\Delta) \le \exp \{ -1/2\}/ \sqrt{2\pi}$ and $\phi(x) \le \frac1{\sqrt{2\pi}}$ for all $x\in \mathbb{R}$ to simplify further.
\begin{equation}
\begin{split}\label{eq:tc-2nd-delta-interim}
        \left| \frac{\partial^2 F^{-1}}{\partial \Delta^2}  \right|
        & \le \frac{\pi\sqrt{1-r^2}}{2\Phi(-\sqrt{2}\Delta)} \left[ \dfrac{2 \exp\left\{ -\frac{1}{2} \right\} }{\sqrt{2\pi}} + \dfrac{\sqrt{2}}{2\pi} + \dfrac{4 \exp\left\{ -\frac{1}{2} \right\}}{\sqrt{2\pi}}    g(r, \Delta) + \dfrac{4}{\sqrt{2\pi}} \left|\frac{\partial g(r, \Delta)}{\partial \Delta} \right| \right]\\
        & \quad + \frac{\pi^2 \sqrt{1-r^2} \left(c_2 + c_1 \sqrt{1-r^2} \right)}{
        2^2\left\{\Phi(-\sqrt{2}\Delta)\right\}^2} \left| \dfrac{2}{2\pi} - \dfrac{4}{\sqrt{2\pi}}  g(r, \Delta) \right|.
\end{split}
\end{equation}
Next, we find the upper bound for $g(r, \Delta)$ and $\frac{\partial g(r, \Delta)}{\partial \Delta}$.
\begin{equation}
\begin{split}\nonumber
    g(r, \Delta) & =\int_{-\infty}^{0} \int_{-\infty}^{0} \phi_2 \left(x_2, x_3 | x_1 = -\Delta; \bSigma_2 (r) \right) dx_2 dx_3\\
    & = \int_{-\infty}^{0} \int_{-\infty}^{0} \phi \left(x_3 | x_2; rx_2^2, 1-r^2 \right) dx_3 \phi \left(x_2; -\Delta/\sqrt{2}, 1/2  \right)dx_2\\
    & = \int_{-\infty}^{0} \int_{-\infty}^{0} \dfrac{1}{\sqrt{2\pi(1-r^2)}} \exp\left\{ -\dfrac{(x_3 - rx_2)^2}{2(1-r^2)}\right\} dx_3  \dfrac{1}{\sqrt{\pi}} \exp\left\{ -\left(x_2 + \frac{\Delta}{\sqrt{2}}\right)^2\right\}  dx_2\\
    & = \dfrac{1}{\sqrt{\pi}} \int_{-\infty}^{0} \Phi\left(\dfrac{-rx_2}{\sqrt{1-r^2}}\right) \exp\left\{ -\left(x_2 + \frac{\Delta}{\sqrt{2}}\right)^2\right\}  dx_2.
\end{split}
\end{equation}
For bivariate random variable $\left(x_2, x_3 | x_1 = -\Delta\right)$ in \eqref{eq:cond-23given1}, the conditional distribution $X_3 | X_2 = x_2$ is normally distributed with mean $rx_2^2$ and variance $1-r^2$, denoted as $\phi \left(x_3 | x_2; rx_2^2, 1-r^2 \right)$ in the second line. $\phi \left(x_2; -\Delta/\sqrt{2}, 1/2  \right)$ denotes probability density function of a normal variable $x_2$ with mean $-\Delta/\sqrt{2}$ and variance $1/2$. The upper bound for $g(r, \Delta)$ is
\begin{equation}
    |g(r, \Delta)| \le \dfrac{1}{\sqrt{\pi}} \int_{-\infty}^{0} \exp\left\{ -\dfrac{1}{2}\left(\dfrac{x_2 + \Delta/\sqrt{2}}{1/\sqrt{2}}\right)^2\right\}  dx_2 = \Phi(\Delta) \le 1.
\end{equation}
The first derivative of $g(r, \Delta)$ with respect to $\Delta$ is
\begin{equation}
\begin{split}\nonumber
    \dfrac{\partial g(r, \Delta)}{\partial \Delta} & = \dfrac{1}{\sqrt{\pi}} \int_{-\infty}^{0} \Phi\left(\dfrac{-rx_2}{\sqrt{1-r^2}}\right) \exp\left\{ -(x_2 + \Delta/\sqrt{2})^2\right\} (-\sqrt{2}x_2 - \Delta)  dx_2\\
    & = -\sqrt{\dfrac{2}{\pi}} \int_{-\infty}^{0} x_2\Phi\left(\dfrac{-rx_2}{\sqrt{1-r^2}}\right) \exp\left\{ -(x_2 + \Delta/\sqrt{2})^2\right\}  dx_2 -\Delta g(r,\Delta),
\end{split}
\end{equation}
and the upper bound is
\begin{equation}
\begin{split}\nonumber
    \left| \dfrac{\partial g(r, \Delta)}{\partial \Delta} \right| \le  \left| \sqrt{\dfrac{2}{\pi}} \int_{-\infty}^{0} x_2 \exp\left\{ -(x_2 + \Delta/\sqrt{2})^2\right\}  dx_2 \right| + |\Delta|\Phi(\Delta).
\end{split}
\end{equation}
Using the change of variable technique via $y = \sqrt{2}(x_2 + \Delta/\sqrt{2})$ yields
\begin{equation}
    \begin{split}\nonumber
    &\left| \sqrt{\dfrac{2}{\pi}} \int_{-\infty}^{0} x_2 \exp\left\{ -(x_2 + \Delta/\sqrt{2})^2\right\}  dx_2 \right|\\
        &  \quad = \left|\dfrac{\sqrt{2}}{\sqrt{2\pi}} \int_{-\infty}^{\Delta} \left(\dfrac{y}{\sqrt{2}} - \dfrac{\Delta}{\sqrt{2}}\right) \exp\left\{ -\dfrac{y^2}{2}\right\}  dy \right|\\
        & \quad = \left|\dfrac{\sqrt{2}}{\sqrt{2\pi}} \int_{-\infty}^{\Delta} \dfrac{y}{\sqrt{2}} \exp\left\{ -\dfrac{y^2}{2}\right\}  dy \right| + \left|\dfrac{\sqrt{2}}{\sqrt{2\pi}} \int_{-\infty}^{\Delta} \dfrac{\Delta}{\sqrt{2}}\exp\left\{ -\dfrac{y^2}{2}\right\}  dy \right|\\
        & \le \dfrac{1}{\sqrt{2\pi}} \exp\left\{ -\dfrac{\Delta^2}{2}\right\} + |\Delta|\Phi(\Delta).
    \end{split}
\end{equation}
Therefore, assume that $\Delta \le M$ for a positive $M$,
\begin{equation}\nonumber
    \left| \dfrac{\partial g(r, \Delta)}{\partial \Delta} \right|
     = \dfrac{1}{\sqrt{2\pi}} \exp\left\{ -\dfrac{\Delta^2}{2}\right\} + 2|\Delta|\Phi(\Delta) \le \dfrac{1}{\sqrt{2\pi}} + 2M.
\end{equation}
Plugging the upper bound of $g(r,\Delta)$ and $\dfrac{\partial g(r, \Delta)}{\partial \Delta}$ into \eqref{eq:tc-2nd-delta-interim} gives
\begin{equation}
\begin{split}\nonumber
        \left| \frac{\partial^2 F^{-1}}{\partial \Delta^2}  \right|
        & \quad \le \frac{\pi\sqrt{1-r^2}}{2\Phi(-\sqrt{2}\Delta)} \left[ c_3 + \dfrac{4}{\sqrt{2\pi}} \left(\dfrac{1}{\sqrt{2\pi}} + 2M\right)\right] + \frac{\pi^2 \sqrt{1-r^2} \left(c_2 + c_1 \sqrt{1-r^2} \right) c_4 }{
        2^2\left\{\Phi(-\sqrt{2}\Delta)\right\}^2}
\end{split}
\end{equation}
where $ c_3 = \dfrac{2 \exp\left\{ -\frac{1}{2} \right\} }{\sqrt{2\pi}} + \dfrac{\sqrt{2}}{2\pi} + \dfrac{4 \exp\left\{ -\frac{1}{2} \right\}}{\sqrt{2\pi}} \approx 1.68$ and $c_4 = \dfrac{4}{\sqrt{2\pi}} -  \dfrac{1}{\pi} \approx 1.28$. Note that $\sqrt{1-r^2} \ge 1-r^2$ for all $r \in (-1, 1)$. To further simplify, we use $\dfrac{\pi}{2} c_3 + 1 \le 4$, $2\sqrt{2\pi} \le 6 $ and  $\dfrac{\pi^2 c_4(c_1+c_2)}{2^2} \le 5$. This concludes the proof.

\hfill \qedsymbol



\begin{lemma}\label{lem:inv_diff1}
Let $f^{-1}$ is an inverse function of $f$ such that $y=f(x)$ and $x = f^{-1}(y)$. Then,
\begin{equation}\nonumber
    \frac{\partial f^{-1}(y)}{\partial y} = \frac1{ \frac{\partial f(x)}{\partial x}}\quad\mbox{(expressed in $y$)}.
\end{equation}
\end{lemma}
\noindent\textbf{\textit{Proof of Lemma~\ref{lem:inv_diff1}}.}  We take a derivative with respect to $x$ on both side of $f^{-1}(y)= x$, then we obtain $\dfrac{\partial f^{-1}(y) }{ \partial  x} = 1$. By the chain rule, 
\begin{equation}\nonumber
        \dfrac{\partial f^{-1}(y)  }{ \partial y}\dfrac{\partial  y }{ \partial x} = 1.
\end{equation}
Therefore, 
\begin{equation}
    \begin{split}\nonumber
        \dfrac{\partial f^{-1}(y)}{y} & =\frac1{\frac{\partial  y }{ \partial x}} =  \dfrac{1 }{ \frac{\partial f(x)}{\partial x} }.
    \end{split}
\end{equation}
\hfill \qedsymbol


\begin{lemma}\label{lem:inv_diff2}
Let $f^{-1}$ be an inverse function of $f$ such that $y=f(x)$ and $x = f^{-1}(y)$. Then,
\begin{equation}\nonumber
    \frac{\partial^2 f^{-1}(y)}{\partial y^2} = -\frac{\partial^2 f(x)}{\partial x^2} \left(\frac1{ \frac{\partial f(x)}{\partial x}}\right)^3 \quad\mbox{(expressed in $y$)}.
\end{equation}
\end{lemma}
\noindent\textbf{\textit{Proof of Lemma~\ref{lem:inv_diff2}}.} From the result of  Lemma~\ref{lem:inv_diff1}, we differentiate $\frac{\partial f^{-1}(y)}{\partial y}$ one more time with respect to $y$ using the chain rule to obtain
\begin{equation}
    \begin{split}\nonumber
        \frac{\partial }{\partial y} \frac{\partial f^{-1}(y)}{\partial y} & = \frac{\partial }{\partial y} \frac1{ \frac{\partial f(x)}{\partial x}} = - \frac1{ \left(\frac{\partial f(x)}{\partial x} \right)^2 } \frac{\partial^2 f(x)}{\partial x^2}   \frac{\partial x}{\partial y}
        = - \frac1{ \left(\frac{\partial f(x)}{\partial x} \right)^2 } \frac{\partial^2 f(x)}{\partial x^2}  \frac{\partial f^{-1}(y)}{\partial y}.
    \end{split}
\end{equation}
\hfill \qedsymbol


\begin{lemma}\label{lem:2dnormal_deriv}
For the bivariate normal probability density function with mean $\mathbf{0}$ and correlation $r$,
\begin{equation}\nonumber
    \phi(x_1, x_2; r) = \frac1{2\pi}\frac1{\sqrt{1-r^2}}\exp\left[-\frac1{2(1-r^2)}\left\{x_1^2 +2r x_1x_2 + x_2^2\right\}\right],
\end{equation}
the partial derivative with respect to $r$ and $x_1$ is
\begin{equation}
\begin{split}\nonumber
    \frac{\partial \phi(x_1, x_2; r)}{\partial r} 
    &= \phi(x_1, x_2; r)\left[\frac{r}{1-r^2} -\frac{x_1x_2}{1-r^2} - \frac{(x_1^2 + 2rx_1x_2 + x_2^2)r}{(1-r^2)^2}\right]\\
    \frac{\partial \phi(x_1, x_2; r)}{\partial x_1}
    & = -\phi(x_1, x_2; r)\dfrac{x_1 + rx_2}{1-r^2}.
\end{split}
\end{equation}
\end{lemma}

\noindent\textbf{\textit{Proof of Lemma~\ref{lem:2dnormal_deriv}}.} 

Consider
\begin{equation}
\begin{split}\nonumber
    \frac{\partial \phi(x_1, x_2; r)}{\partial r} &= \left\{\frac{\partial}{\partial r}\frac1{\sqrt{1-r^2}}\right\} \frac{1}{2\pi}\exp\left[-\frac{x_1^2 +2r x_1x_2 + x_2^2}{2(1-r^2)}\right] \\
    & \quad + \phi(x_1, x_2; r) \left[\frac{\partial}{\partial r} \left\{-\frac{x_1^2 +2r x_1x_2 + x_2^2}{2(1-r^2)}\right\}\right]\\
    & = -\frac1{2}
    \frac1{\sqrt{1-r^2}} \frac{-2r}{1-r^2}\frac1{2\pi}\exp\left[-\frac1{2(1-r^2)}\left\{x_1^2 +2r x_1x_2 + x_2^2\right\}\right] \\
    & \quad  + \phi(x_1, x_2; r) \left\{-\frac{2x_1x_2}{2(1-r^2)} - \frac{x_1^2 +2r x_1x_2 + x_2^2}{2} \left(-\frac{-2r}{(1-r^2)^2}\right)\right\}\\
    &= \phi(x_1, x_2; r)\left[\frac{r}{1-r^2} -\frac{x_1x_2}{1-r^2} - \frac{(x_1^2 + 2rx_1x_2 + x_2^2)r}{(1-r^2)^2}\right].
\end{split}
\end{equation}
Next
\begin{equation}
\begin{split}\nonumber
    \frac{\partial \phi(x_1, x_2; r)}{\partial x_1} &= \phi(x_1, x_2; r)\left[\frac{\partial}{\partial x_1} \left\{-\frac1{2(1-r^2)}\left(x_1^2 +2r x_1x_2 + x_2^2\right) \right\}\right]\\
    &= -\phi(x_1, x_2; r)\frac1{2(1-r^2)}\{2x_1 + 2rx_2\}\\
    & = -\phi(x_1, x_2; r)\dfrac{x_1 + rx_2}{1-r^2}.
\end{split}
\end{equation}
\hfill \qedsymbol

    

\bibliographystyle{biomAbhra}

\bibliography{Bfast}